\documentclass[11pt]{article}
\usepackage{color}
\usepackage[colorlinks=true,citecolor=blue]{hyperref}
\usepackage{amssymb,amsmath,amsfonts,amsthm,amsthm,graphicx}
\usepackage{mathtools}
\usepackage[english]{babel}
\usepackage[utf8]{inputenc}
\usepackage[margin=24mm]{geometry}
\usepackage{times,fourier}

\makeatletter
\renewcommand\section{\@startsection {section}{1}{\z@}%
  {-2ex \@plus -1ex \@minus -.2ex}{1ex \@plus.1ex}%
  {\normalfont\bf\sffamily\color{darkblue}}}
\renewcommand\subsection{\@startsection{subsection}{2}{\z@}%
  {-1.75ex\@plus -0.4ex \@minus -.2ex}{0.6ex \@plus .1ex}%
  {\normalfont\small\bf\sffamily}}
\renewcommand\subsubsection{\@startsection{subsubsection}{3}{\z@}%
  {-0.6ex\@plus -0.2ex \@minus -.2ex}{0.4ex \@plus .1ex}%
  {\normalfont\normalsize\it}}
\renewcommand\paragraph{\@startsection{paragraph}{4}{\z@}%
  {0.2ex \@plus0.2ex \@minus0.1ex}{-0.5em}%
  {\normalfont\normalsize\bfseries}}
\def\ps@headings{%
  \let\@oddfoot\@empty
  \let\@evenfoot\@empty
  \def\@evenhead{\small\sffamily\thepage\hfil\slshape\leftmark}%
  \def\@oddhead{\small\sffamily{\slshape\rightmark}\hfil\thepage}%
  \let\@mkboth\markboth
  \def\chaptermark##1{\markboth{{\ifnum \c@secnumdepth >\m@ne
		\if@mainmatter \@chapapp\ \thechapter. \ \fi \fi ##1}}{}}%
  \def\sectionmark##1{\markright {{\ifnum \c@secnumdepth >\z@
		\thesection. \ \fi ##1}}}}
\def\fbf#1{\setbox0=\hbox{$#1$}\kern-0.10\wd0
  \lower0.02em\copy0\kern-\wd0 \lower0.02em\hbox{\kern+0.04em\copy0}\kern-\wd0
  \raise0.00em\copy0\kern-\wd0 \raise0.00em\hbox{\kern-0.04em\box0}}
\def\overl@ss#1#2{\vcenter{\offinterlineskip
        \ialign{$\m@th#1\hfil##\hfil$\crcr#2\crcr<\crcr } }}
\def\gl{\mathrel{\mathpalette\overl@ss>}}
\graphicspath{{figures/}}
\makeatother

\advance\textwidth -6mm
\advance\hoffset 4mm
\advance\textheight -6mm
\advance\voffset 0mm
\advance\parskip 2pt
\numberwithin{equation}{section}
1

\def\maketitle{\par\noindent{\LARGE\bf\sffamily\thetitle}\\[1.4ex]
{\large\theauthor}\\[0.6ex]
\textit{\theaddress}\\[0.2ex]
{\small\today}\par\vglue1.4\bigskipamount}
\def\title#1{\def\thetitle{#1}}
\def\author#1{\def\theauthor{#1}}
\def\address#1{\def\theaddress{#1}}

\def\be{\begin{equation}}
\def\ee{\end{equation}}
\def\bse{\begin{subequations}}
\def\ese{\end{subequations}}

\definecolor{deeppurple}{rgb}{0.5, 0, 0.7}
\definecolor{darkblue}{rgb}{0, 0, 0.7}

\def\sech{\mathop{\rm sech}\nolimits}
\def\diag{\mathop{\rm diag}\nolimits}

\def\dn{\mathop{\rm dn}\nolimits}

\def\Integer{\mathbb{Z}}
\def\Real{\mathbb{R}}
\def\Complex{\mathbb{C}}

\def\i{\text{i}}

\def\Re{\mathop{\rm Re}\nolimits}
\def\Im{\mathop{\rm Im}\nolimits}

\def\tr{\mathop{\rm tr}\nolimits}
\def\re{\mathrm{re}}
\def\im{\mathrm{im}}
\def\d{\mathrm{d}}
\def\e{\mathrm{e}}

\def\@#1{{\mathbf{#1}}}
\def\_#1{{\mathsf{#1}}}
\let\~=\tilde
\let\==\bar
\let\^=\hat
\let\<=\langle
\let\>=\rangle

\def\min{\mathop{\rm min}\nolimits}
\def\max{\mathop{\rm max}\nolimits}

\def\note[#1]{\marginpar{\color{blue}[#1]}}

\def\half{{\textstyle\frac12}}

\def\C{{\mathbb C}}

\def\F{{\mathcal F}}

\def\z{\zeta}
\def\l{\lambda}
\makeatother

\let\trueparagraph=\paragraph
\def\paragraph#1{\par\smallskip\trueparagraph{\rm\textbf{#1}}}
\allowdisplaybreaks

\def\bse{\begin{subequations}}
\def\ese{\end{subequations}}

\begin{document}
\pagestyle{plain}

\title{\color{darkblue}
Semiclassical dynamics and coherent soliton ensembles in the derivative 
nonlinear Schrödinger equation with periodic initial conditions}
\author{Zachery Wolski$^1$, Zechuan Zhang$^2$, Gino Biondini$^3$ and Gregor Kova\v{c}i\v{c}$^1$}
\address{$^1$ Department of Mathematical Sciences, Rensselaer Polytechnic Institute, Troy, NY 12180 - USA\\ 
$^2$ Scuola Internazionale Superiore di Studi Avanzati, Trieste, 34136 - Italy\\ 
$^3$ Department of Mathematics, State University of New York at Buffalo, Buffalo, NY 14260 - USA}
\maketitle

\begin{quotation}
\small\noindent
\small
\textbf{Abstract.}~
The semiclassical limit of the derivative nonlinear Schrodinger equation with periodic initial conditions is studied analytically and numerically. The spectrum of the associated scattering problem for a certain class of initial conditions, referred to as periodic single-lobe potentials, is numerically computed, and it is shown that the spectrum becomes confined to the real and imaginary axes or the spectral parameter in the semiclassical limit. A formal Wentzel-Kramers-Brillouin expansion is computed for the scattering eigenfunctions, which allows one to obtain asymptotic expressions for the  number, location and size of the spectral bands and gaps.  The results of these calculations  suggest that, in the semiclassical limit, all excitations in the spectrum become effective solitons.
Finally, the analytical predictions are compared with direct numerical simulations as well as with numerical calculations of the Lax spectrum, and the results are shown to be in excellent agreement.
\end{quotation}
\par\medskip

\section{Introduction}

The derivative nonlinear Schrödinger (DNLS) equation, 
\be
\label{e:DNLS}
\i q_t+ q_{xx}+\i(|q|^2q)_x=0, 
\ee
where subscripts denote partial differentiation, 
appears in a variety of physical contexts, including the modeling of circularly polarized Alfvén waves in plasmas and aspects of turbulence and nonlinear optics \cite{Mj1976, Mj1978, Mj1986, Mj1989}. 

As was shown by Kaup and Newell \cite{KN1978}, the DNLS equation \eqref{e:DNLS} is the compatibility condition of the overdetermined pair of linear systems
\bse
\begin{gather}
    \phi_x = U\,\phi\,,\qquad
    U(x,t,\zeta) = -\i\zeta^2\sigma_3+\zeta Q\,,
    \label{e:KN}\\
    \phi_t = V\,\phi\,,\qquad
    V(x,t,\zeta) = -2\i\z^4\sigma_3+2\z^3Q - \i\z^2|q|^2\sigma_3+\z|q|^2Q - \i\z Q_x\sigma_3\,,
\end{gather}
\ese
where $\phi:\Real\times\Real^+\to\C^2$ is assumed to be a $C^2$ complex-valued function of $x$ and $t$, $\z\in\C$ is the $(x,t)$-independent spectral parameter, and $Q$ is the $(x,t)$-dependent matrix potential given by
\begin{equation}
    Q(x,t)=\begin{bmatrix}
        0 & q(x,t)\\
        -q^*(x,t) & 0
    \end{bmatrix}\,,
\end{equation}
where $\sigma_3=\diag(1,-1)$ is the third Pauli matrix and the asterisk denotes complex conjugation. The linear equation \eqref{e:KN} is usually referred to as the Kaup-Newell spectral problem.

As is standard, the Lax pair formulation of \eqref{e:DNLS} enables solution of the initial-value problem via the inverse scattering transform (IST). The IST for \eqref{e:DNLS} with localized potentials (i.e., zero boundary conditions at infinity) was developed in \cite{KN1978} and later revisited in \cite{HuangChen90,Steudel2003,ZhouHuang2007,Zhou2012} to construct multi-soliton solutions by different approaches. The IST with nonzero boundary conditions (NZBCs) was first considered by Kawata and Inoue \cite{KawataInoue78}, yielding the associated one-soliton solutions. Subsequently, two-soliton solutions for both zero and nonzero boundary conditions were obtained by Kawata \textit{et al.} \cite{Kawata+1979}. Chen and Lam later investigated the non-decaying setting using Riemann–Hilbert method \cite{ChenLam04}, and subsequently multi-soliton solutions under NZBCs were studied in \cite{ChenYangLam06}. Moreover, Liu, Perry, and Sulem applied the IST to the well-posedness of the DNLS equation \cite{LiuPerrySulem}, and Pelinovsky \textit{et alii} later addressed well-posedness without the small-norm assumption in the IST framework, both with and without solitons \cite{PSS2017,PS2018}. Most recently, Killip \textit{et al.} established well-posedness for the DNLS in $L^2(\Real)$~\cite{Killip2024}.

Following the Its-Matveev formula \cite{ItsMatveev75},  Its and Matveev first obtained the finite-gap solutions for the DNLS equation \cite{ItsMatveev83}.
Chow and Ng \cite{ChowNg2011} analyzed the simplest periodic solutions of DNLS via the polar decomposition, and  quasi-periodic (finite-gap) solutions to the DNLS equation have been studied using the complex finite-dimensional Hamiltonian systems in \cite{ChenZhang2020, geng+2015, Wright2020, ZhaoFan2020}.
Recently, Pelinovsky and Chen \cite{Chen+2021} classified all periodic standing waves of DNLS and characterized their spectral stability with respect to localized perturbations, and characterize the modulational instability of the periodic standing waves in the DNLS equation. They further show that the space-time localization of a rogue wave is only
possible if the periodic standing wave is modulationally unstable \cite{ChenPelinovsky2021}.

A complementary viewpoint is the semiclassical limit, where a small parameter $\epsilon$ scales dispersion and $\epsilon\downarrow 0$. For the gauge-equivalent modified NLS equation, DiFranco and Miller developed an IST-based modulation framework and a spectral description of the semiclassical scattering data \cite{Miller2008}. DiFranco, Miller, and Muite further identified subsonic, supersonic, and transsonic sectors and the associated phase transitions \cite{Miller2011}, providing a template for the DNLS equation.
The semiclassical behavior of solutions of \eqref{e:DNLS} with periodic boundary conditions has not been explored much in the literature to the best of our knowledge.
In this work we aim to address this issue.  
Specifically, we study the DNLS equation in the small dispersion limit and we characterize analytically and numerically the behavior of the Lax spectrum and derive asymptotic expansions for the number of bands, band widths and gap widths.

The rest of this paper is organized as follows.  
In section~\ref{s:ICs} we present the problem set-up and the results of numerical simulations of the DNLS equation that illustrate the dynamical behavior we are interested in studying.
In section~\ref{s:spectrum} we present the results of numerical calculations of the IST spectrum illustrating its behavior in the semiclassical limit. 
In section~\ref{s:asymptotics} we present the set-up of the WKB expansion and the main results on the characterization of the spectrum.
In section~\ref{s:numerics} we describe the methods used to numerically integrate the DNLS equation in time and numerically compute the Lax spectrum.
In section~\ref{s:wkb} we present the details of the asymptotic calculations.
Finally, in section~\ref{s:conclusions} we end this work with some concluding remarks.

\section{Semiclassical DNLS Equation with Single-Lobe Periodic Potentials}
\label{s:ICs}

The semiclassical limit, or small dispersion limit, of integrable dispersive nonlinear wave equations has been a classical topic 
of research since the pioneering works of Lax and Levermore on the Korteweg-de\,Vries (KdV) equation 
\cite{LaxLevermore1,LaxLevermore2,LaxLevermore3}.
In recent years, the topic has been receiving renewed interest due on one hand to the rich mathematical structure of solutions and on the other hand to its relevance to a variety of physical phenomena such as dispersive shock waves and soliton gases (e.g., see \cite{ElHoefer2016,JStatMech2021p114001,PRE109p061001} and references therein).
The DNLS equation in a semiclassical scaling is
\be
\label{e:dnls-eps}
\i q_t+\epsilon q_{xx}+\i(|q|^2q)_x=0,
\ee
where the factor $\epsilon$ can be introduced via a simultaneous rescaling of both $x$ and~$t$
in~\eqref{e:DNLS}. 
One then supplements~\eqref{e:dnls-eps} with initial conditions (ICs) that are independent of~$\epsilon$, and 
looks at the problem of determining the behavior of solutions as a function of~$\epsilon$.
In practice, this corresponds to the problem of looking at situations in which the ICs vary over spatial scales that are much longer than the typical dispersion length of the PDE.

\subsection{Initial Conditions}

Below we study the dynamics of solutions of \eqref{e:dnls-eps} generated by a class of initial conditions referred to as ``single-lobe periodic potentials'', as in \cite{BO2020}.
These are defined as real-valued functions $q(x,0)$ such that $q(x+2L,0) = q(x,0)$ $\forall x\in\Real$, 
and which are increasing for $x\in(-L,x_{\text{max}})$ and decreasing for $x\in(x_{\text{max}},L)$, with $x_{\text{max}}\in(-L,L)$ being the location where the maximum value of $q(x,0)$ is achieved. 
For simplicity, we will also assume that $q(x,0)$ is even, as this condition will simplify the calculations of the asymptotic behavior of the spectrum, although it is not strictly necessary for the analysis.

In particular, for concreteness we will consider three specific potentials as examples, similarly to~\cite{BO2020}: 
\vspace*{-1.4ex}
\begin{subequations}
\label{ics}
\begin{gather}
        q_{\cos}(x,0)=(\cos(x)+1)/2\,,
        \label{raisedcos}\\
        q_{\exp\sin}(x,0)=\e^{-\sin^2(x)}\,,
        \label{expsin}\\
        q_{\dn}(x,0)=\dn(x|m),\qquad m\in[0,1]\,,
        \label{dn}
\end{gather}
\end{subequations}
where $\dn(\,\cdot\,)$ is one of the Jacobi elliptic functions \cite{NIST} and $m$ is the elliptic parameter.
We emphasize, however, that the theoretical analysis that will be presented in section~\ref{s:asymptotics} applies to any real-valued periodic single-lobe potentials.
Hereafer, the potential in equation \eqref{raisedcos} will be referred to as the ``raised cosine" potential, 
the one in equation~\eqref{expsin} will be referred to as the ``exp sine" potential, 
and the one in equation~\eqref{dn} as the ``dn" potential.  
Recall that the period of the latter is $2K(m)$, where $K(\cdot)$ is the complete elliptic integral of the first kind~\cite{NIST},
and at the extremal values of $m$, one has $\dn(x|0)=1$ and $\dn(x|1)=\sech x$.

\subsection{Dynamical Behavior, Dispersionless System and Dispersive Shock Regularization}

\begin{figure}[t!]
\centerline{\includegraphics[scale=.565]{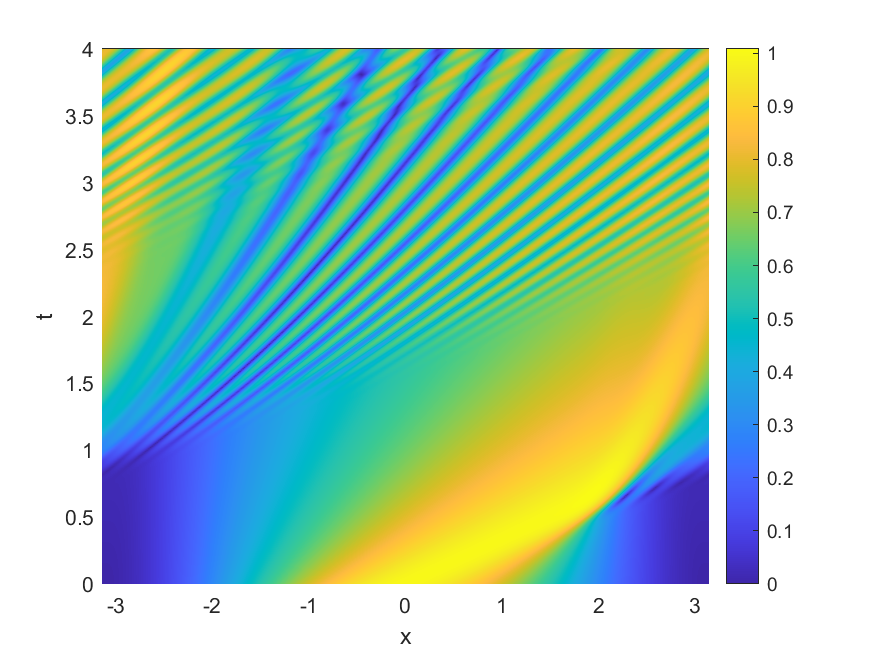}\includegraphics[scale=.565]{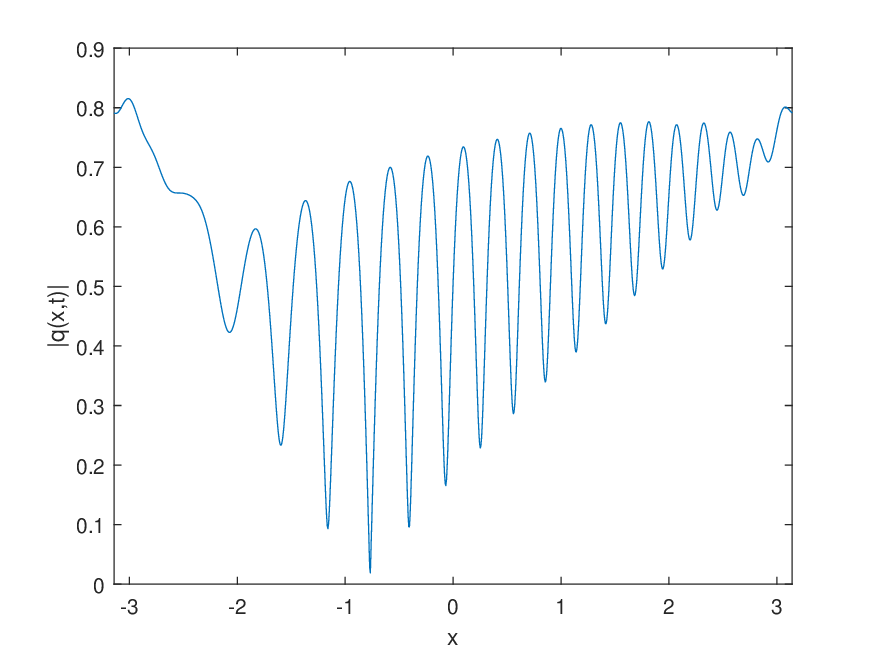}}
\centerline{\includegraphics[scale=.565]{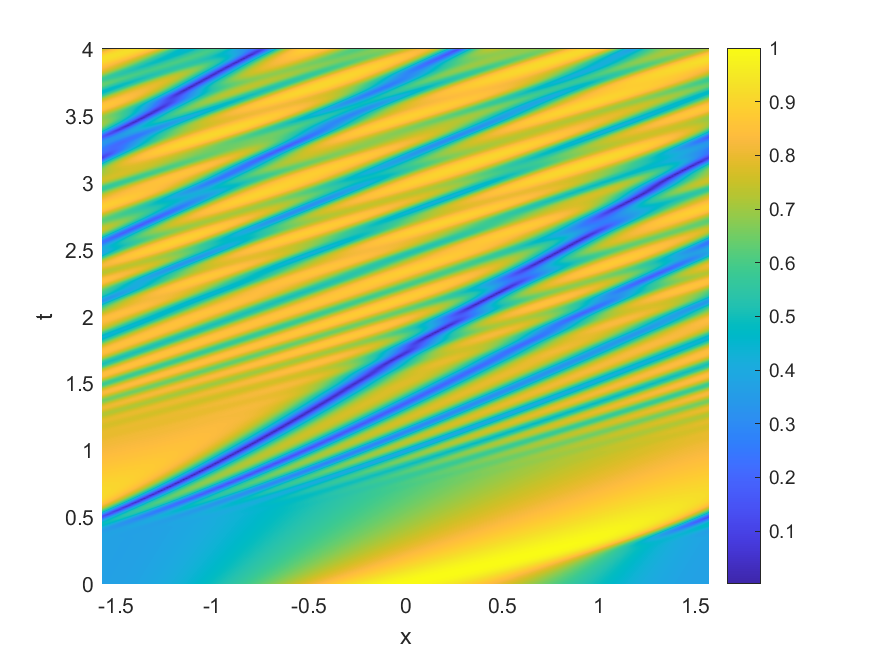}\includegraphics[scale=.565]{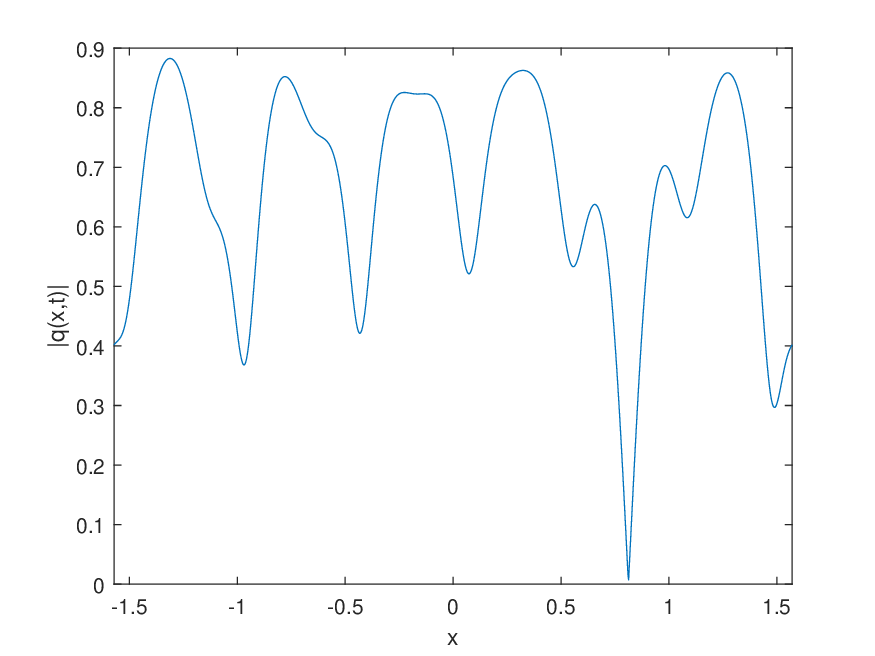}}
\centerline{\includegraphics[scale=.565]{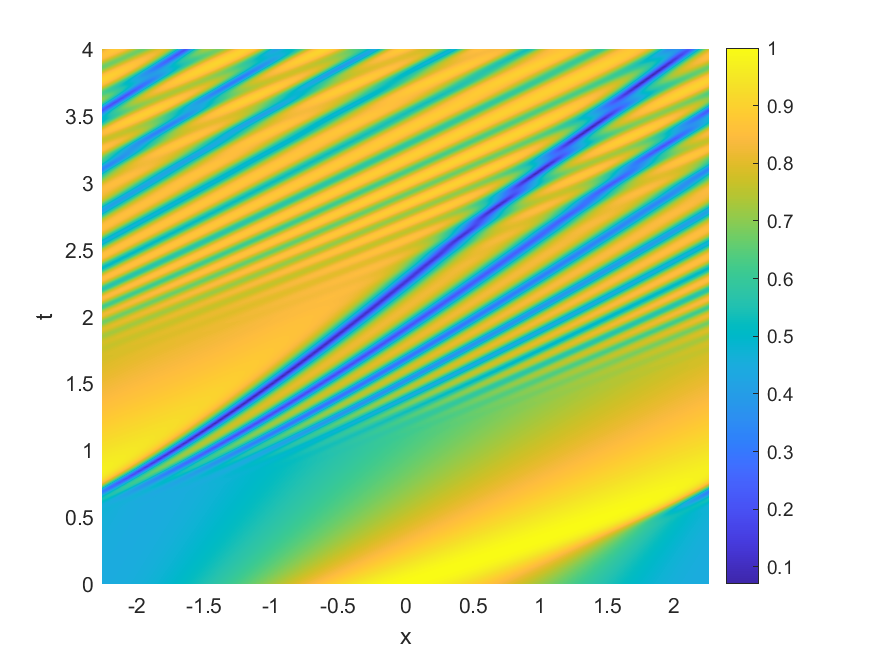}\includegraphics[scale=.565]{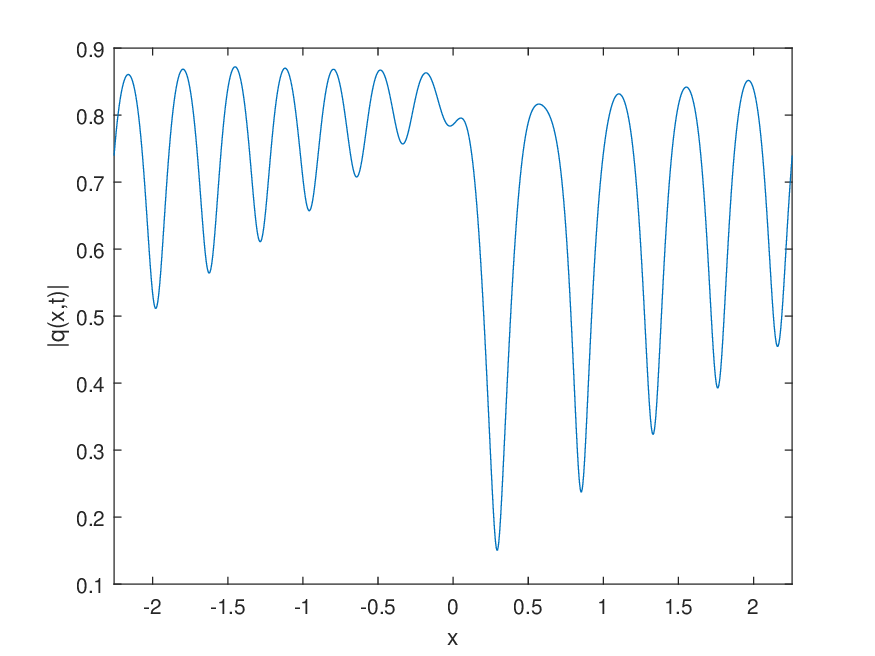} }
\caption{Left column: Density plots of the amplitude $|q(x,t)|$ of the solution of the DNLS equation in the semi-classical limit with different initial conditions and $\epsilon = 0.05$. 
Right column: Corresponding snapshots of the solution at $t=2.5$.
Top row: the ``raised cosine" from equation \eqref{raisedcos}. 
Middle row: the ``exp sine" potential from equation \eqref{expsin}. 
Bottom row: the ``dn" potential from equation \eqref{dn} with elliptic parameter $m=0.8$.
}
\label{fig1}
\end{figure}

We numerically integrated equation~\eqref{e:dnls-eps} with initial conditions given by~\eqref{ics} using a pseudo-spectral method in space and fourth-order method in time (see section~\ref{s:numerics} for details). 
The left column of figure~\ref{fig1} shows density plots of the raised cosine, exp sine, and dn initial conditions for a specific value of $\epsilon$ and, in the case of the dn potential, of the elliptic parameter~$m$.
The right column of figure~\ref{fig1} shows the corresponding profiles of $|q(x,t)|$ as a function of~$x$
at $t=2.5$.

All the plots in figure~\ref{fig1} exhibit self-steepening and shock-like behavior at short times. 
This should not be too surprising, as
the formal limit $\epsilon\to0$ of equation~\eqref{e:dnls-eps} yields
\vspace*{-1ex}
\begin{equation}
    q_t+(|q|^2q)_x = 0\,.
\label{e:dnlsdispersionless}
\end{equation}
Equation~\eqref{e:dnlsdispersionless} is a first-order system of quasi-linear equations for the real and imaginary parts of the solution.  Indeed, letting 
$q_\re(x,t) = \Re(q)$, $q_\im(x,t) = \Im(q)$ and $\@q(x,t) = (q_\re,q_\im)^T$, 
we can rewrite equation~\eqref{e:dnlsdispersionless} as the system
\vspace*{-1ex}
\begin{equation}
\@q_t + A(\@q)\,\@q_x = 0\,,\qquad
A(\@q) = \begin{bmatrix} 
   3q_\re^2 + q_\im^2 & 2q_\re q_\im \\
   2 q_\re q_\im & 3q_\im^2 + q_\re^2
  \end{bmatrix}\,.
\end{equation}
The characteristic speeds of the system are the eigenvalues of $A(\@q)$, which are 
$q_\re^2 + q_\im^2$ and $3(q_\re^2 + q_\im^2)$.
Therefore, both the real and imaginary portions of the solution move in the positive~$x$ direction.
Moreover, regions where $|q|^2$ is large move faster than regions where $|q|^2$ is small, 
leading eventually to a gradient catastrophe and shock formation, 
as the faster moving portions of the solution catch up to the slower moving portions. 

When $\epsilon\ne0$, the solution of the full DNLS equation~\eqref{e:dnls-eps} can be expected to be close to
that of the dispersionless system~\eqref{e:dnlsdispersionless} as long as the latter is smooth and slowly varying.
As it often happens with dispersive nonlinear equations, however,
the dispersive term causes oscillations to develop near the point of the gradient catastrophe.
Similarly to what happens in the KdV equation \cite{ZabuskyKruskal}, 
these oscillatory excitations 
then propagate and eventually interact with each other due to the periodicity of the boundary conditions.
The following sections are aimed at quantitatively characterizing the spectrum of these nonlinear excitations.

To understand the differences in the behaviors of the plots in figure~\ref{fig1}, one can  
begin by noting that the maximum of each initial condition is~1 
(which can always be made the case without loss of generality thanks to the scaling invariance of the DNLS equation) and assume the differences in $\epsilon$ are not significant enough to affect the behaviors of the plots. The raised cosine initial condition having a minimum of 0 has the most abrupt shock. The exp sine intial condition has a minimum of approximately 0.37 and the dn initial condition with $m=0.8$ has a minimum of approximately 0.45 and the profile of their shocks are relatively similar to each other as their minima are similar. The dn initial condition with $m=0.2$ has a minimum of approximately 0.89. The smaller difference between the maximum and minimum for this initial condition could explain why it takes longer for the shock dynamics to develop and why they are less pronounced.

In passing, it is also worthwhile to note that there are many solutions of equation~\eqref{e:dnls-eps} that do not form a shock. 
For example, equation~\eqref{e:dnls-eps} admits plane wave solutions of the form
$q(x,t)=A\,\e^{\i[kx+(\epsilon k^2+A^2k)t]}$, with $A>0$ and $k$ an arbitrary real constant.
These solutions do not form a shock, as $|q|^2=A^2$ is constant. 
Note, however, that these are not single-lobe periodic potentials, nor they are real-valued, and will therefore not be discussed in the remainder of this paper.
At the same time, we should note that some of these solutions are modulationally unstable, and would therefore give rise to a soliton gas under appropriate circumstances
(e.g., see \cite{Zhong2025} for a recent study).

\section{Spectrum of the DNLS Equation in the Semiclassical Limit}
\label{s:spectrum}

In this section we present the basic theoretical framework that will be used to investigate the Lax spectrum of the DNLS equation with periodic boundary conditions in the semiclassical limit.

\subsection{Lax Pair and Monodromy Matrix}

The semiclassical DNLS equation \eqref{e:dnls-eps} is the compatibility condition of the semiclassical Lax pair
\begin{subequations}
\label{e:laxpair}
\begin{gather}
    \label{scattering}
    \epsilon\phi_x = U_\epsilon \phi\\
    \epsilon\phi_t = V_\epsilon \phi\,,
\end{gather}
\end{subequations}
with
\begin{subequations}
\label{e:Laxmatrices}
    \begin{gather}
        \label{e:firstLax}
        U_\epsilon(x,t,\zeta)=-\i\zeta^2\sigma_3+\zeta Q\\
        V_\epsilon(x,t,\zeta)=-2\i\zeta^4\sigma_3+2\zeta^3Q-\i\zeta^2Q^2\sigma_3 + \zeta\left(Q^3 - \i\epsilon Q_x\sigma_3\right)\,.
    \end{gather}
\end{subequations}
The parameter $\zeta$ appearing in~\eqref{e:Laxmatrices} is referred to as the scattering (or spectral) parameter. 
The first half of the Lax pair~\eqref{e:firstLax} is similar to the first half of the Lax pair for the focusing NLS equation, with the key difference being that the entire right hand side of~\eqref{e:firstLax} is multiplied by an extra factor of $\zeta$. 
In the case of the focusing NLS equation, the first half of the Lax pair can be rewritten as the eigenvalue problem for a one-dimensional Dirac operator, in which case $\zeta$ and $\phi(x,t,\zeta)$ are the eigenvalues and corresponding eigenfunctions of the Dirac operator. 
In the case of the DNLS equation, the extra factor of~$\zeta$ creates a quadratic eigenvalue problem. 
Nevertheless, like with the NLS equation, the Lax spectrum $\Sigma$ is defined as the set of values of $\zeta\in\Complex$ for which \eqref{scattering} admits solutions
$\phi(x,t,\zeta)$ bounded over all $x\in\Real$.

If the potential $q(x)$ in \eqref{e:firstLax} is $2L$-periodic, Floquet-Bloch theory (e.g., see \cite{Eastham})
implies that all bounded solutions have the form
\begin{equation}
    \phi(x,\zeta)=\e^{\i\nu x}w(x,\zeta)
\end{equation}
where $w(x,\zeta)=w(x+2L,\zeta),$ $i\nu$ is referred to as the Floquet exponent, and $\nu\in[0,\pi/L)$. 
(Here and below, the time dependency has been suppressed for brevity, 
since the IST spectrum is time-independent.)
Furthermore, the Floquet multipliers $\mu=\e^{2\i\nu L}$ are the eigenvalues of the monodromy matrix $M(\zeta)$ defined as
\begin{equation}
\label{e:monodromy}
    M(\zeta)=\Phi(x-L,\zeta)^{-1}\Phi(x+L,\zeta)
\end{equation}
where $\Phi(x,\zeta)$ is any fundamental solution matrix of \eqref{e:firstLax}. 
As the right hand side of \eqref{e:firstLax} is traceless, $\det\Phi\equiv1$ and $\det M\equiv1$, 
implying that the Floquet multipliers $\mu$ are the roots of the polynomial
\begin{equation}
    \mu^2-2\Delta(\zeta)\,\mu+1=0\,,
\end{equation}
where $\Delta(\zeta) = \frac12 \tr M(\zeta)$ is the Floquet discriminant.
In order for~\eqref{scattering} to have bounded solutions, one then needs $|\mu|=1$, which in turn requires 
$\Im\Delta(\zeta)=0$ and $\Re\Delta(\zeta)\in[-1,1]$. 
The Floquet-Bloch spectrum can be defined as 
\begin{equation}
    \label{e:fb_specturm}
    \Sigma_\nu = \left\{ \zeta\in\Complex:~\Delta(\zeta) = \cos(2\nu L) \right\}.
\end{equation}
The Lax spectrum is then the union of all Floquet-Bloch spectra, namely
\be
\Sigma=\cup_{\nu\in[0,\pi/L)}\Sigma_\nu\,.
\label{e:LaxFloquetunion}
\ee
Like with the NLS equation, the Lax spectrum is invariant under time evolution. 
Accordingly, even though reality and evenness are not preserved under time evolution, 
the semiclassical limit of the spectrum of the periodic single-lobe potentials is still an interesting object of study, since it is shared by a much broader class of solutions of the DNLS equation.

\subsection{Behavior of the Lax Spectrum in the Semiclassical Limit}

\begin{figure}[t!]
\centerline{\includegraphics[scale=.5]{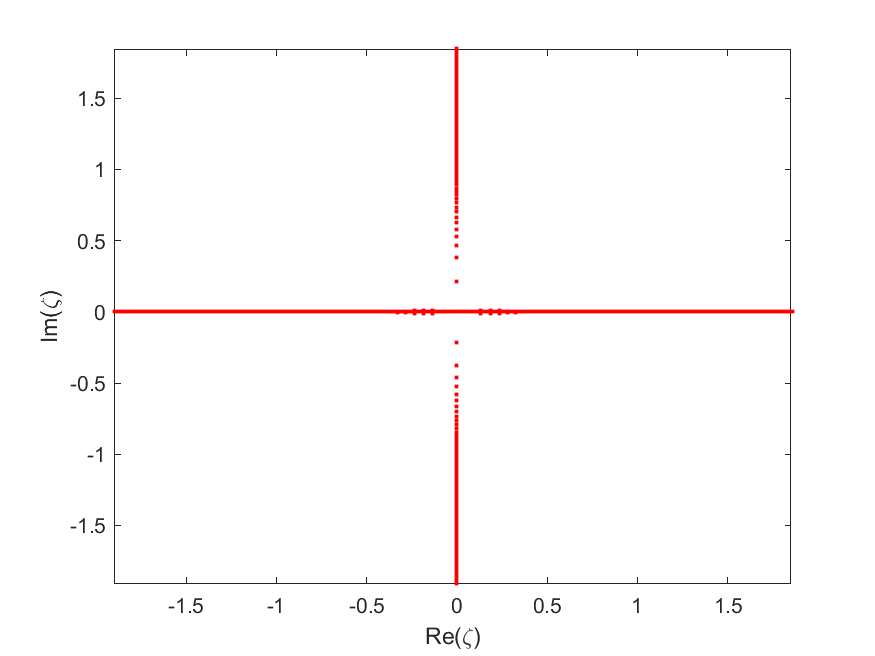}
  \includegraphics[scale=.5]{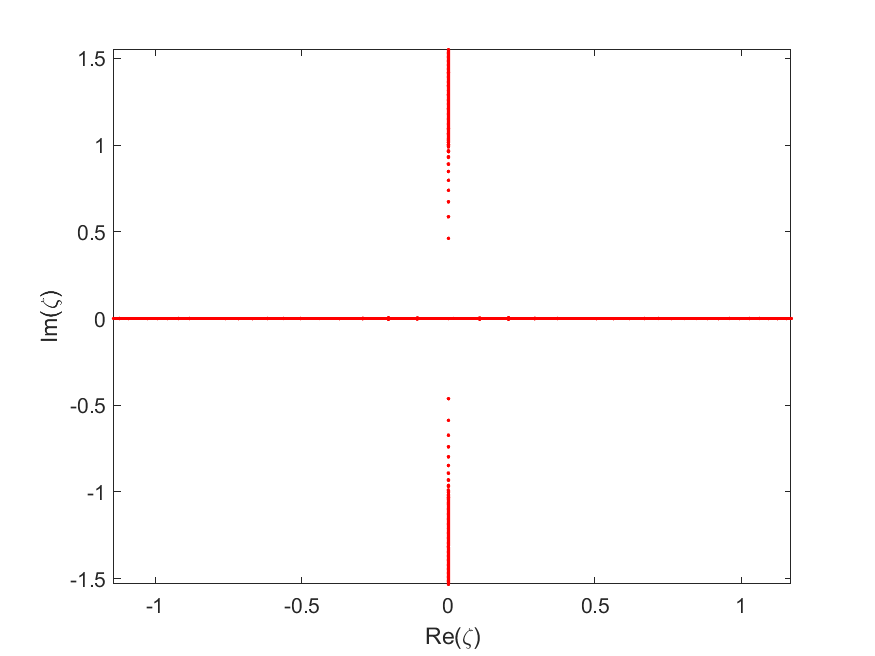}}
\centerline{\includegraphics[scale=.5]{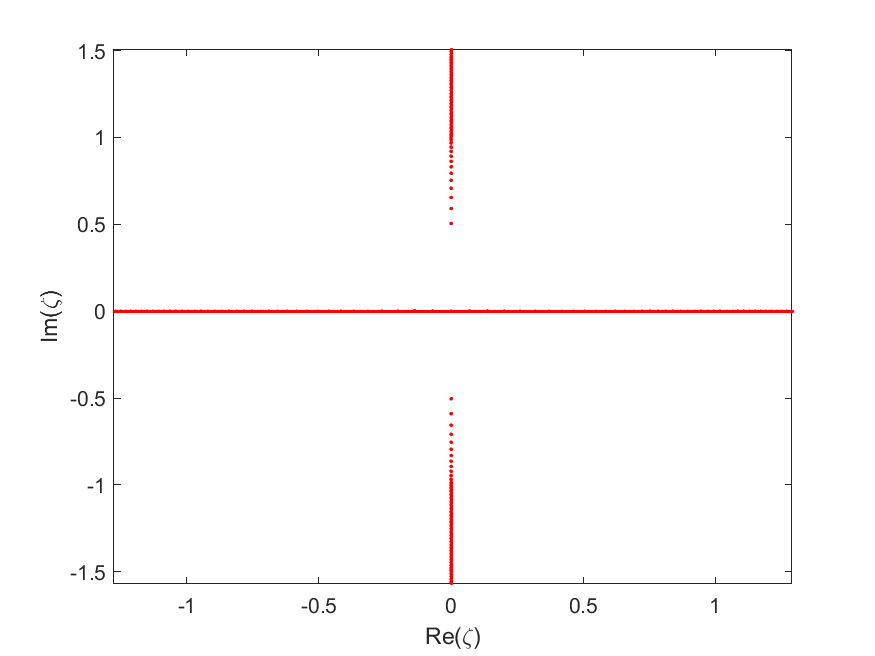}
  \raise2ex\hbox{\includegraphics[scale=.5]{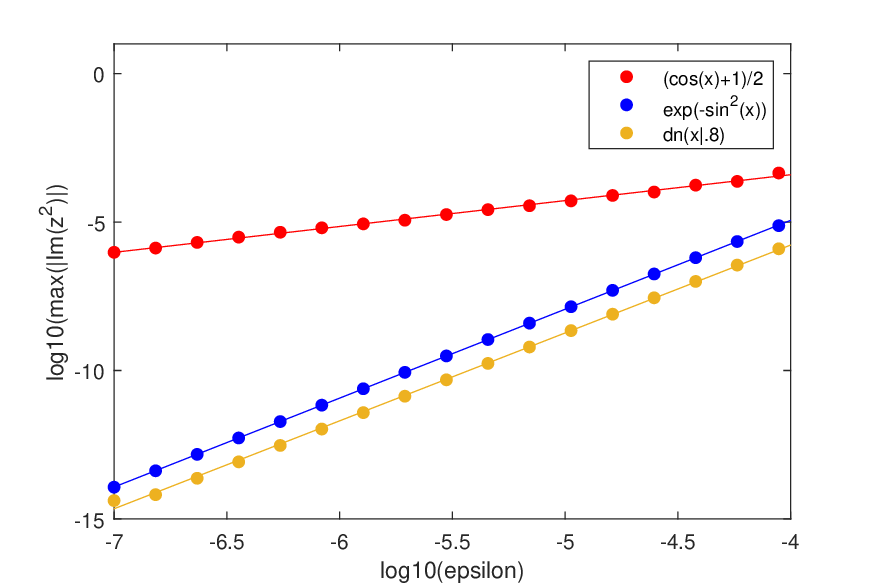}}}
\caption{Plots of the numerically computed Floquet spectrum for the three initial conditions in equations~\eqref{ics}, plus a plot showing the convergence of the spectrum to the real axis as $\epsilon\downarrow0$. 
Top Left: the raised cosine initial condition with $\epsilon=0.07$. 
Top Right: the exp sine initial condition with $\epsilon=0.07.$. 
Bottom left: the dn initial condition with $m=0.8$ and $\epsilon=0.07$. 
Bottom Right: convergence to the real axis as $\epsilon\downarrow0$. The way deviation from real axis is measured is the magnitude of Im$(\zeta^2)$. 
Circles: numerically calculated maximum of Im$(\zeta^2)$ over all points in the Lax spectrum as a function of $\epsilon$, for a range of values of $\epsilon$ from $\epsilon=10^{-4}$ to $\epsilon=10^{-7}$. 
Solid lines: linear regressions of these points in a loglog space.}
\label{fig2}
\end{figure}

We now show that in the semiclassical limit $\epsilon\downarrow0$ the Lax spectrum simplifies significantly. Using a modified Floquet-Hill's method that has been used to great success examining the Lax spectrum of the nonlinear Schrodinger equation (e.g., see \cite{PRE2025,BO2020} and references therein), 
we numerically calculated the spectrum of the first half of the Lax pair \eqref{e:firstLax} with a variety of single-lobe periodic potentials for different values of~$\epsilon$. 
Some of the corresponding results are shown in figure~\ref{fig2}.
These plots clearly illustrate remarkable similarities between all three of the single-lobe periodic potentials in equation~\eqref{ics}. 
Namely: (i) the entire real axis appears to be an uninterrupted band of Lax spectrum (spectral band), 
(ii) there appears to be spectral bands extending uninterrupted past a point to both positive and negative infinity along the imaginary axis, and 
(iii) the only other pieces of spectrum lie on the imaginary axis. 

The most intriguing property, however, is that all of the Lax spectrum appears to localize to the real and imaginary axis in the semiclassical limit. 
We can define the deviation from the real and imaginary axis as the maximum of the absolute value of the imaginary portion of $\zeta^2$, i.e., $d = \max(|\Im(\zeta^2)|)$.
The bottom right panel of figure~\ref{fig2} shows the value of $d$ in the Lax spectrum corresponding to a variety of values $\epsilon$ in the range $(10^{-4},10^{-7})$, 
One can clearly see that $d$ exhibits power-law behavior as a function of~$\epsilon$, 
i.e., $d=O(\epsilon^\alpha)$ as $\epsilon\downarrow0$
In particular, a linear regression fit of the numerically computed values yields
$\alpha=0.872\pm0.664$ for the raised cosine potential, $\alpha=2.994\pm2.278$ for the exp sine potential, and $\alpha=2.961\pm2.254$ for dn$(x,0.8)$, with the intervals representing the 99\% confidence bands around the slope of the linear regression fit.

The fact that the Lax spectrum converges to the real and imaginary axis in the semiclassical limit has the practical consequence that that, by setting $\lambda=\zeta^2$ and observing the symmetries in \eqref{e:firstLax}, the entire spectrum can effectively be viewed as function of the real $\lambda$ axis, a property that will prove extremely useful in the asymptotic analysis of the spectrum, discussed in the next section.

\section{Asymptotic Analysis of the Scattering Problem and Effective Solitons}
\label{s:asymptotics}

In this section we present the WKB analysis that is used to study the behavior of the Lax spectrum and derive asymptotic expansions for the number of bands and the relative band widths.
For brevity, the detailed calculations are presented in section~\ref{s:wkb}, and only the main results are presented here. 

\subsection{Asymptotic Analysis of the Scattering Problem}

\begin{figure}[t!]
\centerline{\includegraphics[scale=.5]{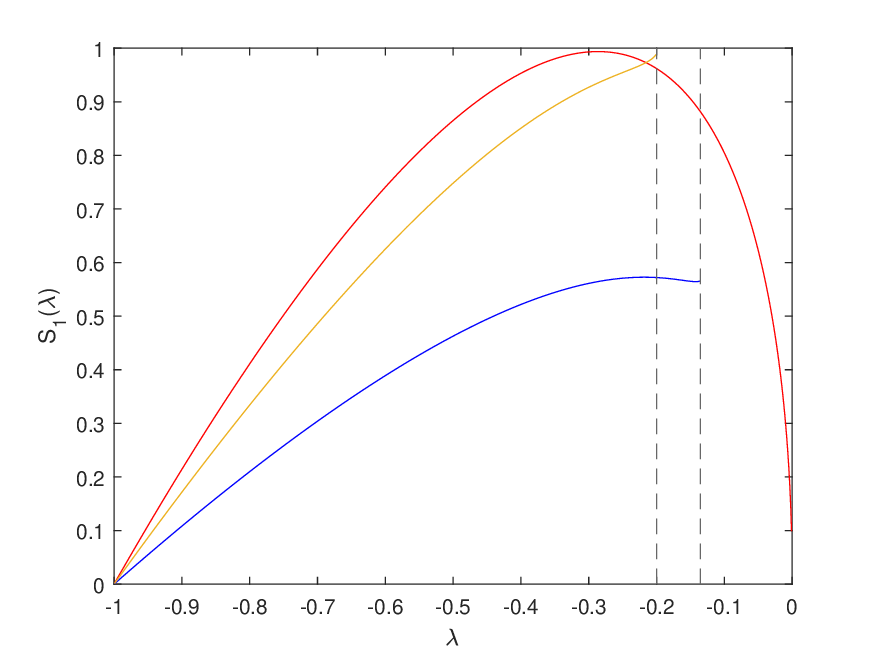}
  \includegraphics[scale=.5]{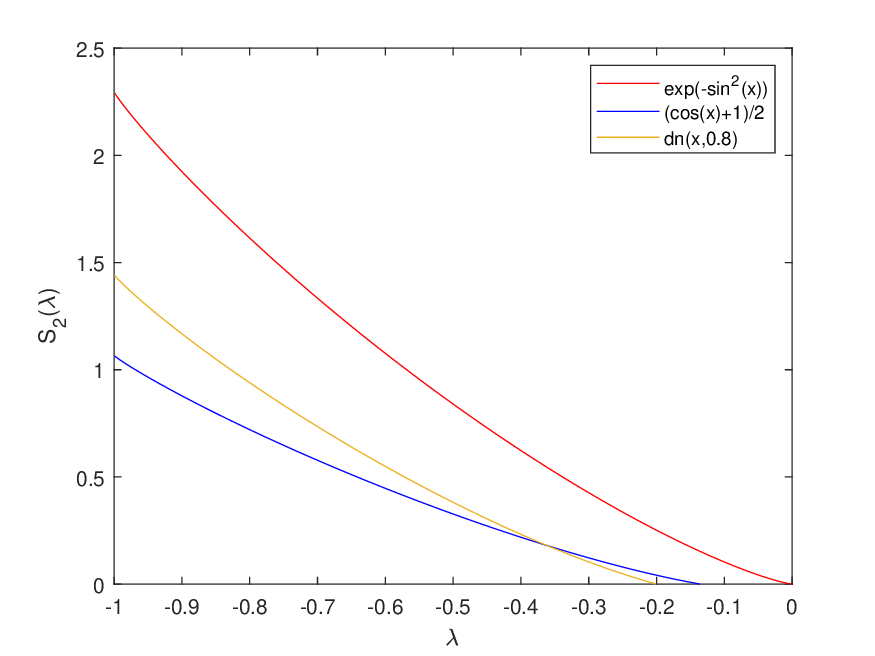}}
\caption{Plots of $S_1(\lambda)$ (left) and $S_2(\lambda)$ (right). One of the dashed lines in the $S_1(\lambda)$ plot is at $-e^{-2}$, the upper $\lambda$ boundary of range (iii) for the $\exp(-\sin^2(x))$ potential. The other dashed line corresponds to the upper boundary of range (iii) for the dn potential with $m=0.8.$}
\label{fig3}
\end{figure}

The change of variable $v=\phi_1+\i\phi_2$ and $v^*=\phi_1-\i\phi_2$ maps equations \eqref{e:firstLax} with a real potential to
\begin{equation}
\label{e:vODE}
    \epsilon^2v''+(\i\epsilon\sqrt{\lambda}q'(x)+Z(x,\lambda))v=0
\end{equation}
with
\begin{equation}
    Z(x,\lambda)=\lambda(\lambda-q^2(x)),\qquad \lambda=\zeta^2.
\end{equation}
This formulation suggests the use of the WKB method to obtain an asymptotic description of the Lax spectrum. 
When the potential $q(x)$ in \eqref{e:firstLax} is single-lobe, the real $\lambda$ axis splits into several disjoint regions, determined by the possible signs that $Z(x,\lambda)$ can attain and by whether or not the minimum value the potential, $q_\mathrm{min}$, is zero.
Specifically:
\begin{enumerate}
    
    \item[(i)] 
    For $\lambda\in(-\infty,-q^2_{\text{max}})$, we have that $Z(x,\lambda)>0$ for all $x\in\Real$, and there are no turning points in the WKB analysis. 
    The WKB expansion with $\lambda$ in this range yields the following asymptotic expression for the Floquet discriminant:
    \vspace*{-1ex}
    \begin{equation}
        \Delta(\lambda) = \cos\left(\int_{-L}^L\sqrt{Z(x,\lambda)}dx/\epsilon\right).
    \label{e:Deltacosine}
    \end{equation}
    The resulting values of $\Delta(\lambda)$ are always real and in $[-1,1]$. 
    As a result, all values $\lambda\in(-\infty,-q_{\text{max}}^2)$ are part of the Lax spectrum.
    
    \item[(ii)] 
    For $\lambda\in(0,\infty)$, we again have that $Z(x,\lambda)>0$ for all $x\in\Real$, and the WKB expansion in this range is once again given by equation~\eqref{e:Deltacosine}.
    As a result, all values $\lambda\in(0,\infty)$ are also always part of the Lax spectrum.
    
    \item[(iii)] 
    For $\lambda\in(-q^2_{\text{max}},-q_{\text{min}}^2)$, we have that $Z(x,\lambda)$ has two symmetric turning points at $x = \pm p(\lambda)$, which are the $\lambda$-dependent values of $x$ at which $Z(\pm p(\lambda),\lambda)=0$. 
    To handle the turning points, the interval $(-L,L)$ must be broken into subregions, different representations of the eigenfunctions must be made for each of the subregions, and the coefficients of these eigenfunctions must be asymptotically matched. 
    The result of this process yields the Floquet discriminant in this region as 
    \vspace*{-1ex}
    \begin{equation}
    \label{e:trM3}
            \Delta(\lambda) = \cos(2S_2(\lambda)/\epsilon)\cosh(S_1(\lambda)/\epsilon+\ln(2))
    \end{equation}
    with
    \vspace*{-1ex}
    \begin{equation}
        S_1(\lambda)=\int_{-p(\lambda)}^{p(\lambda)}\sqrt{|Z(x,\lambda)|}dx,\quad
        S_2(\lambda)=\int_{p(\lambda)}^L\sqrt{|Z(x,\lambda)|}dx.
    \label{e:S1S2def}
    \end{equation}
    Thus, in this case both the frequency and the amplitude of the oscillations of $\Delta(\lambda)$ increase as $\epsilon\downarrow0$. 
    As a result, the length of the spectral bands in this range will shrink as $\epsilon$ decreases due to the increases in amplitude, but the number of bands will increase due to the increase in frequency.
    A detailed analysis of the properties of bands and gaps in this region is performed in section~\ref{s:bands}.

    Plots of $S_1(\lambda)$ and $S_2(\lambda)$ for the three examples in~\eqref{ics}
    are shown in figure~\ref{fig3}.
    We point out that, comparing the above analysis to that of the single-lobe potentials for the focusing NLS in the semiclassical limit \cite{BO2020}, the resulting expressions for $\tr M(\lambda)$ in the region with turning points are similar, 
    with $S_1(\lambda)$ and $S_2(\lambda)$ having the same definitions as in \cite{BO2020}, 
    but their role is reversed compared to \cite{BO2020}.
    That is, for the focusing NLS equation, $S_2(\lambda)$ appears as in the argument of the hyperbolic cosine, and $S_1(\lambda)$ appears in the argument of the trigonometric cosine.
    
    \item[(iv)] 
    For $\lambda\in(-q_{\text{min}}^2,0)$, we have that $Z(x,\lambda)<0$ $\forall x\in\Real$, and as a result there are no turning points in the domain. 
    The WKB expansion yields the Floquet discriminant as
    \vspace*{-1ex}
    \begin{equation}
        \Delta(\lambda) = \cosh\left(\int_{-L}^L\sqrt{|Z(x,\lambda)|}dx/\epsilon\right).
    \label{e:trM4}
    \end{equation}
    The resulting value is always strictly greater than 1. 
    Therefore the range $\lambda\in(-q_{\text{min}}^2,0)$ is never a part of the Lax spectrum.
    Note however that this range may or may not be non-existent depending on the particular potential considered.

    \item[(v)] 
    A final value of interest is $\lambda=0$. 
    This value does not fall into any of the previous ranges and it is difficult to analyze from a WKB perspective, as $Z(x,\lambda)=0$ for all $x\in\Real$ when $\lambda=0$. 
    However, note that $\lambda=0$ means $\zeta=0$, which, when inserted into equation~\eqref{scattering}, 
    implies that any constant matrix is a fundamental matrix solution solution of the scattering problem. 
    Thus, from equation~\eqref{e:monodromy} we see that the monodromy matrix for $\lambda=0$ is simply the identity matrix regardless of the potential.  Therefore, the value $\lambda=0$ is always a part of the Lax spectrum.

\end{enumerate}

In terms of the original spectral variable $\zeta$, we can summarize the above WKB results by saying that the Lax spectrum contains an infinitely long band consisting of the entire real axis, a generally infinite number of bands along the half lines $(-\i\infty,-\i q_{\max}^2)$ and $(\i q_{\max}^2,\i\infty)$ of the imaginary axis, and a collection of bands and gaps along $(-\i q_{\max}^2,-\i q_{\min}^2)$ and $(\i q_{\min}^2,\i q_{\max}^2)$. 
Moreover, for potentials for which $q_{\min}\neq0$, the portions $(-q_{\min}^2,0)$ and $(0,q_{\min}^2)$  of the imaginary axis will always be gaps.

At this point, it is instructive to compare the Lax spectrum of a single-lobe periodic potential in the semiclassical limit to the spectrum on the real line with nonvanishing boundary conditions \cite{Zhou2012}. 
The spectrum of the problem with nonvanishing boundary conditions on the real line comprises a continuous spectrum, which is the same for all solutions with the same value of $q(x)=q_0$ as $x\rightarrow\pm\infty$, and a discrete spectrum of points in the complex $\zeta$ plane, each of which contributes a solitonic component in the solution. 
The continuous spectrum consists of the entire real axis and the entire imaginary axis, with the exception of gaps from $(-\i q_0,0)$ and $(0,\i q_0)$. 
Any discrete spectrum that is present in those gaps correspond to solitonic components in the solution. 
The bands that appear in the periodic case are not discrete spectrum.
However, as epsilon decreases, the number of bands present in the gap increases, while their width shrinks. 
In the next section we will show that as $\epsilon\downarrow0$, the band widths become sufficiently small that the bands can be effectively treated as discrete spectrum.

\begin{figure}[t!]
\centerline{\includegraphics[scale=.5]{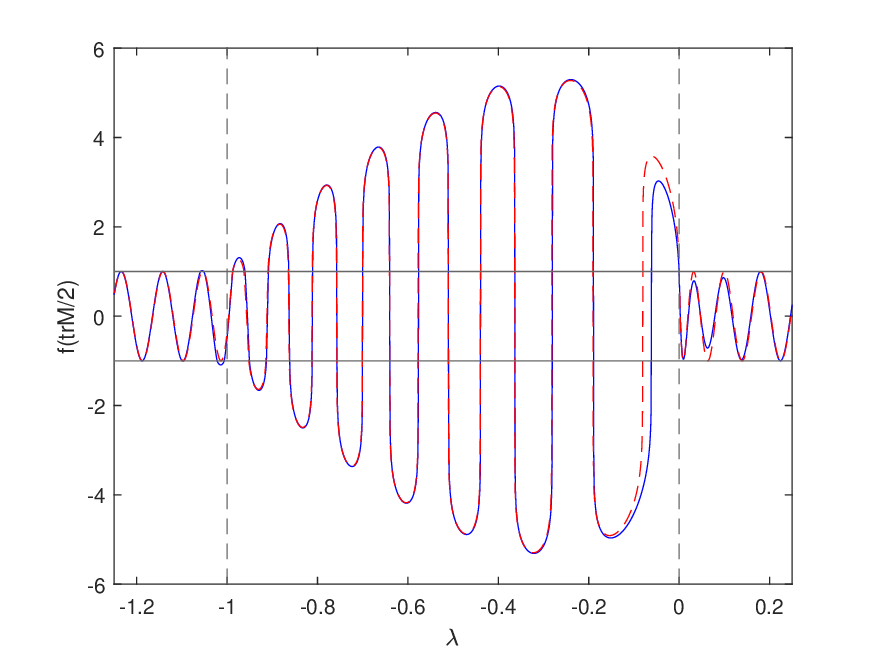}
  \includegraphics[scale=.5]{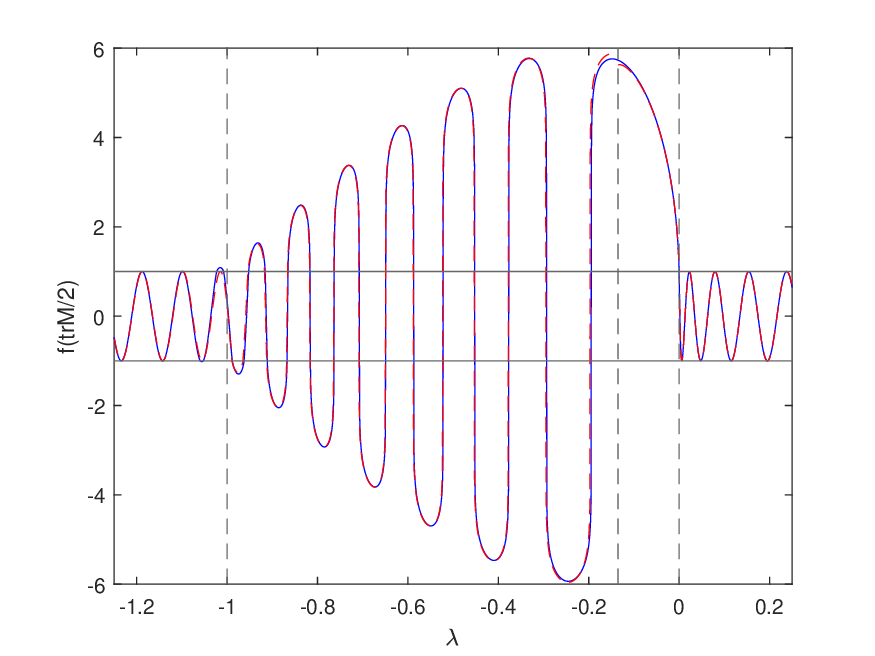}}
\centerline{\includegraphics[scale=.5]{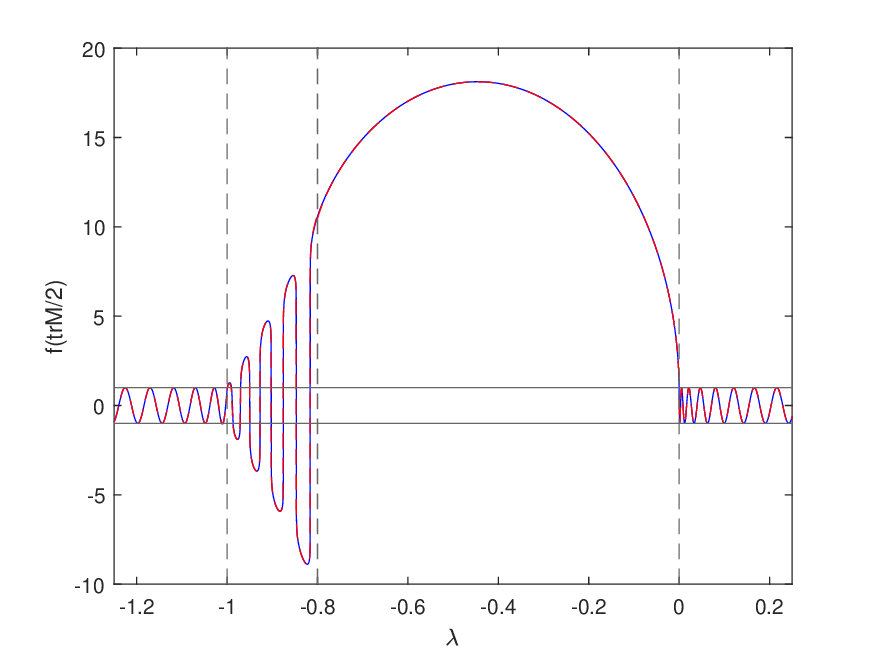}
  \includegraphics[scale=.5]{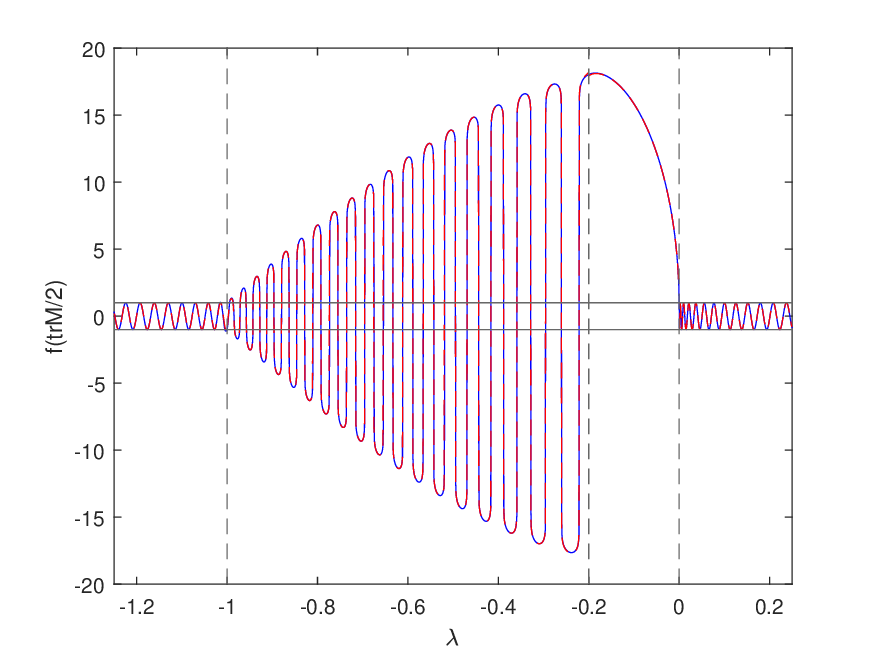}}
\caption{Comparison between the numerically computed Floquet discriminant $\Delta = \tr M/2$ and the WKB approximation. 
Top left: raised cosine potential with $\epsilon=0.1$. 
Top right: exp sine potential with $\epsilon=0.05$. 
Bottom left: dn potential with $m=0.2$ and $\epsilon=0.037$. 
Bottom right: dn potential with $m=0.8$ and $\epsilon=0.025$. 
Red dashed lines: WKB approximation of $\tr M$ as a function $\lambda$. 
Blue solid lines: results of numerically integrating the first half of the Lax pair \eqref{e:firstLax}. 
Vertical dashed lines: the four ranges of $\lambda$, ordered from left to right as (i), (iii), (iv) and~(ii). 
Note that the raised cosine does not have a range~(iv). 
Horizontal lines: the values $\tr M=\pm2$, corresponding to the edges of the spectrum. 
Since the amplitude of the oscillations grows exponentially, following \cite{OsborneBergamasco,BO2020} 
the function $f(\Delta)$ was plotted instead of $\Delta$, in order to capture the whole behavior in a single plot, with $f(y)$ defined as $f(y)=y$ for $|y|\leq1$ and $f(y)=\text{sgn}(y)(1+\log_{10}|y|)$ for $|y|>1$.}
\label{fig4}
\end{figure}

\subsection{Number of Bands, Band Widths and Gap Widths}
\label{s:bands}

Equation~\eqref{e:trM3} clearly implies that the oscillation amplitude grows exponentially as $\epsilon\downarrow 0$
when $\lambda\in(-q^2_{\text{max}},-q_{\text{min}}^2)$. 
This means that each spectral band is narrowly concentrated around one of the zeros of the trace. 
Denoting by $z_n$ the zeros of $\Delta(\lambda)$ in the range $\lambda\in(-q^2_{\text{max}},-q_{\text{min}}^2)$,
we have that the number of such zeros is also the number of spectral bands in this range.
From~\eqref{e:trM3} we see that the zeros $z_n$ are given by the equation
\be\label{e:S2zn}
S_2(z_n)=\half(n-\half)\pi\epsilon,\qquad z_n\in(-q^2_{\max},-q^2_{\min}).
\ee
Since $S_2(\l)$ is a monotonically decreasing function, one therefore obtains
that the number $N_\epsilon$ of such zeros is
\be
\label{e:bandcount}
N_\epsilon=\left\lfloor \frac{2S_2(-q^2_{\max})}{\pi\epsilon}+\frac{1}{2} \right\rfloor\,.
\ee
Next, let $\lambda_n$ for $n=1,2,\dots,2N_\epsilon$ be the decreasing values of $\lambda$ so that $\Delta(\lambda) =\pm 1$.
These values are the edges of the spectral bands, and $\lambda_{2n} < z_n < \lambda_{2n-1}$.
Specifically, $\lambda_{4n}$ and $\lambda_{4n-3}$ are the values such that $\Delta(\lambda)=1$, 
and $\lambda_{4n-2}$ and $\lambda_{4n-1}$ are the values such that $\Delta(\lambda) =-1$. 
Therefore, the $n$-th spectral band is the interval $[\lambda_{2n},\lambda_{2n-1}]$.
Thus, the width of the $n$-th spectral band and that of the $n$-th spectral gap are given by
\be
\label{e:Taylortau}
w_n=\lambda_{2n-1}-\lambda_{2n},\qquad g_n=\lambda_{2n-2}-\lambda_{2n-1}\,,
\ee
Taking a Taylor expansion of the Floquet discriminant $\Delta(\lambda)$ about $\lambda = z_n$ and differentiating yields
\be
\label{e:TaylorDelta}
\Delta(\lambda) = \Delta'(z_n)(\l-z_n)+\half\Delta''(z_n)(\l-z_n)^2+O\left((\l-z_n)^3\right)\,, \quad \l\to z_n\,,
\ee
together with
\bse
\begin{align}
&\Delta'|_{\l=z_n}= \mp\frac{2S_2'(z_n)}{\epsilon}\cosh \left(\frac{S_1(z_n)}{\epsilon}+\ln 2\right)(1+o(1))\,,   \\
&\Delta''|_{\l=z_n}=\frac{1}{\epsilon^2}\e^{S_1(z_n)/\epsilon}(1+o(1))\,.
\end{align}
\ese
Evaluating equation~\eqref{e:TaylorDelta} at $\l_{2n}$, we obtain the following asymptotic expansion as $\epsilon\downarrow 0$:
\be
\l_{2n}-z_n= \left.\frac{1}{\Delta'}\right|_{\l=\l_{2n}}+O\left(\epsilon\e^{-{S_1(z_n)/\epsilon}}\right)\,.
\ee
Thus, as $\epsilon\downarrow 0$, we obtain the following asymptotic expansion for $w_n$:
\be
\label{e:wn}
w_n=\frac{\epsilon}{|S'_2(z_n)|}\sech\left(\frac{S_1(z_n)}{\epsilon}+\ln 2\right)+O\left(\epsilon\e^{-{S_1(z_n)}/{\epsilon}}\right)\,.
\ee
We also have 
\be
(w_n+g_n)-(z_{n+1}-z_n)=O\left(\epsilon\e^{-{S_1(z_n)}/{\epsilon}}\right)\,,
\ee
where we used the fact that
\bse
\begin{gather}
w_n+g_n=(\l_{2n-1}-z_n)+(z_n-\l_{2n})+(\l_{2n-2}-\l_{2n-1}),\\
z_{n+1}-z_n=(\l_{2n-1}-z_n)+(\l_{2n-2}-\l_{2n-1})+(z_{n+1}-\l_{2n-2})\,.
\end{gather}
\ese
From equation~\eqref{e:S2zn}, we directly have $S_2(z_{n+1})-S_2(z_n)=\pi\epsilon/2$. 
Expanding again $S_2(\l)$ about $\lambda = z_n$, evaluating at $\l=z_{n+1}$ and solving for $z_{n+1}-z_n$, 
we have, as $\epsilon\downarrow 0$,
\be
z_{n+1}-\lambda_n=\frac{\pi\epsilon}{2|S_2'(z_{n})|}+O(\epsilon^2)\,.
\ee
Finally, combining all of the asymptotic expressions  above, we can show that
\be
w_n+g_n=\frac{\pi\epsilon}{2|S_2'(z_{n})|}+O(\epsilon^2)\,,\quad \epsilon\downarrow 0\,.
\ee
Introducing the relative band width as
\be
W_n=\frac{w_n}{w_n+g_n}\,,
\ee
we then obtain the following estimate:
\begin{equation}
\label{e:relativewidth}
    W_n=\frac{2}{\pi}\sech\left(\frac{S_1(z_n)}{\epsilon}+\ln2\right).
\end{equation}

\begin{figure}[t!]
\centerline{\includegraphics[scale=.5]{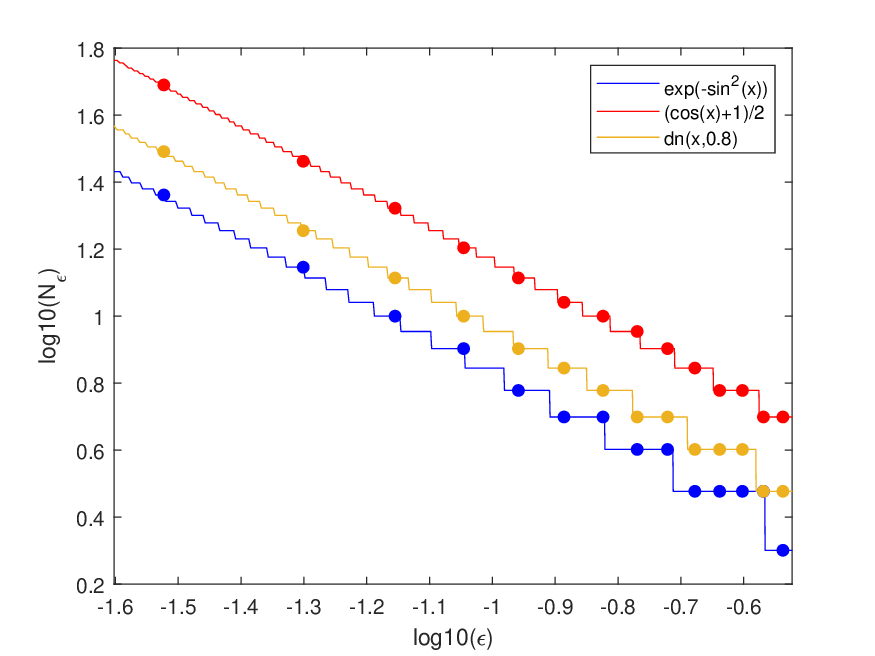}
  \includegraphics[scale=.5]{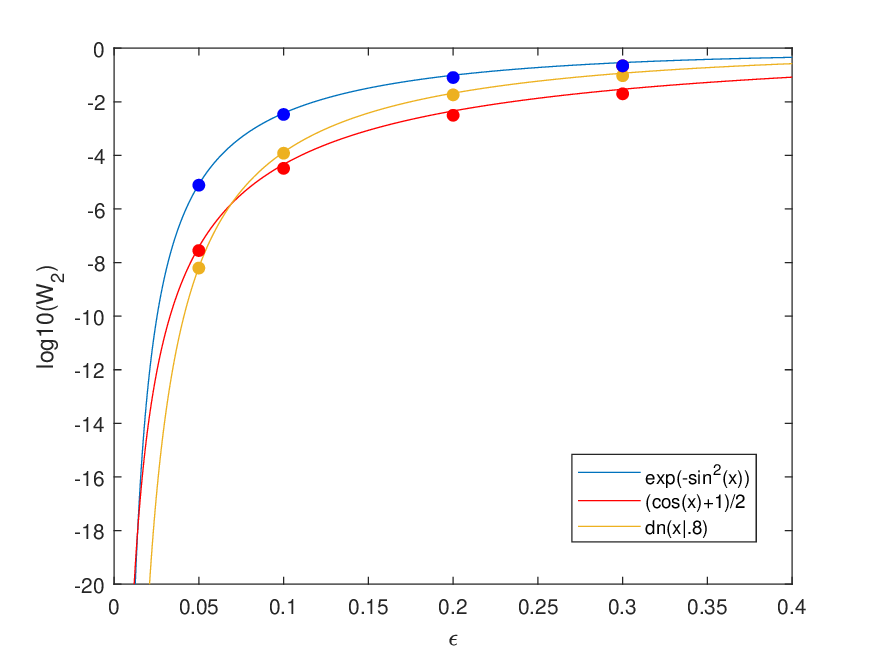}}
    \caption{Left: the number of bands in the range $(-q^2_\mathrm{max},-q^2_\mathrm{min})$ as predicted by equation~\eqref{e:bandcount} (solid lines) as a function of $\epsilon$ and the numerically computed number of bands from the monodromy matrix (circles). 
    Right: the relative band width of the second spectral band $W_2$ as predicted by equation \eqref{e:relativewidth} as solid lines for the three examined potentials. The solid dots are numerically computed values for relative width $W_2$ calculated using band edges generated with Floquet-Hill's method.}
    \label{fig5}
\end{figure}

Figure~\ref{fig5} shows how the analytical predictions in equations~\eqref{e:bandcount} and \eqref{e:relativewidth} 
for $N_\epsilon$ and $W_n$ compare to the numerically calculated values for the three potentials in equation~\eqref{ics}. 
We point out that, for the purposes of this comparison, we choose to look at $W_2$ instead of $W_1$.
The reason for doing so stems from the fact $W_1$ needs to be defined slightly differently for the three potentials.
Recall that he $n$-th spectral gap is defined as $(\lambda_{2n-1},\lambda_{2n-2})$ for $\lambda$ in range~(iii). 
For potentials that have a range~(iv), like the exp sine and dn potential
[i.e., potentials for which $q_\mathrm{min}>0$], 
the value~$\lambda_0$ that should be the right band edge of the first spectral gap is not in range~(iii). 
Accordingly, if one wants an approximation of $W_1$ in these cases, 
one would have to take into account the fact that the entirety of range~(iv) is part of the first gap.
However, the above asymptotic approximations hold for every other value of~$n$, and 
equation~\eqref{e:wn} is valid for $w_n$ showing that the band length decreases as $\epsilon\downarrow0$.

Equation~\eqref{e:relativewidth} clearly shows that $W_n\downarrow0$ as $\epsilon\downarrow0$. 
Similarly to \cite{BO2020}, we can introduce a fixed threshold $\kappa\ll1$, 
and we can declare that any band with a relative band with $W_n<\kappa$ can be effectively thought to be a discrete eigenvalue, and can be considered as such in comparisons with the case of nonvanishing boundary conditions on the real line. 
We next show that: (a)~the choice of $\kappa$ does not matter, and 
(b)~all bands become effective solitons in the limit $\epsilon\downarrow0$.

First, notice from equation \eqref{e:relativewidth} that, for a fixed $\epsilon$, a larger value of $S_1(\lambda)$ results in a smaller relative relative band width $W_n$. 
Next, from the plot of $S_1(\lambda)$ on the left panel of figure~\ref{fig3} we see that, for each potential, there exists a single maximum for $S_1(\lambda)$. 
Accordingly, let $\lambda_L$ and $\lambda_R$ be two values such that $\lambda_L$ and $\lambda_R$ lie respectively to the left and the right of the maximum of $S_1(\lambda)$. 
Moreover, once a value for $\lambda_L$ is selected, the value of $\lambda_R$ is chosen such that $S_1(\lambda_R)=S_1(\lambda_L)$, 
or alternatively $\lambda_R=-q_{\text{min}}^2$ if the equality $S_1(\lambda_R)=S_1(\lambda_L)$ cannot be satisfied. 
Then, setting $\lambda_L=\lambda_n$ such that $W_n<\kappa$, we see that all bands with $\lambda_n$ in the range $(\lambda_L,\lambda_R)$ 
can be effectively treated as discrete eigenvalues, since their relative band widths will be less than $\kappa$.
An implicit expression for $\lambda_L$ can then be written as
\begin{equation}
    S_1(\lambda_L)=\frac{\epsilon}{2}\ln\left(\frac{4}{\pi\kappa}\right).
\end{equation}
It is now clear that $S_1(\lambda_L)$ must approach 0 as $\epsilon\downarrow0$. 
Examining once again the left panel of figure~\ref{fig3}, it can be seen that $\lambda_L$ approaches $-q_{\text{max}}^2$ as $S_1(\lambda_L)$ approaches~zero. 
Then, either $\lambda_R$ will approach $-q_{\text{min}}^2$ (as in the case of the raised cosine potential) or will already be defined as $\lambda_R=-q_{\text{min}}^2$ (as in the case of the exp sine and dn potentials). 
Accordingly, as $\epsilon\downarrow0$ the range $(\lambda_L,\lambda_R)$ will approach the entire range where equation \eqref{e:trM3} is defined, and all bands will become effectively discrete points, regardless of the choice of $\kappa$.

\begin{figure}[b!]
\centerline{\includegraphics[scale=.55]{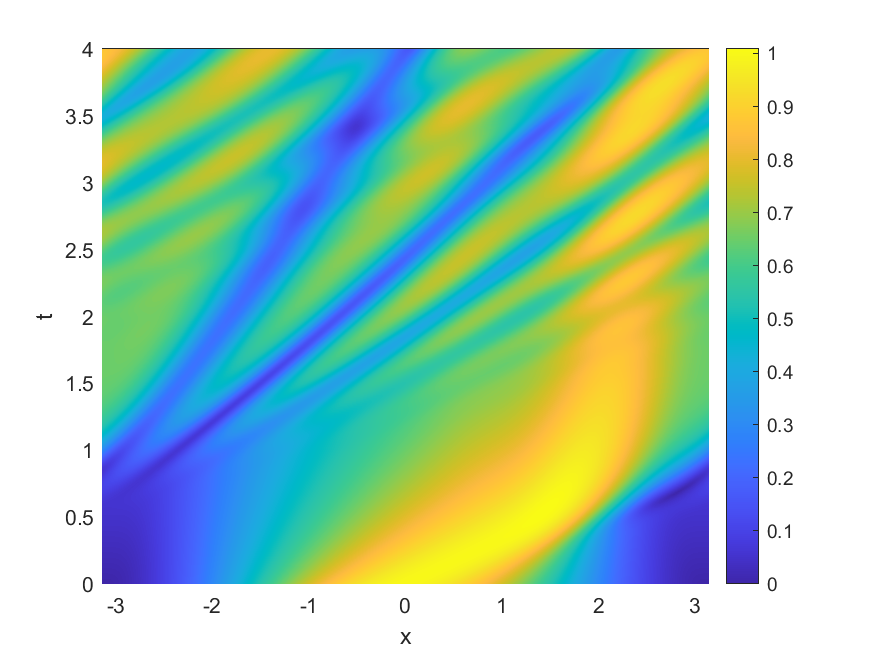}
  \includegraphics[scale=.55]{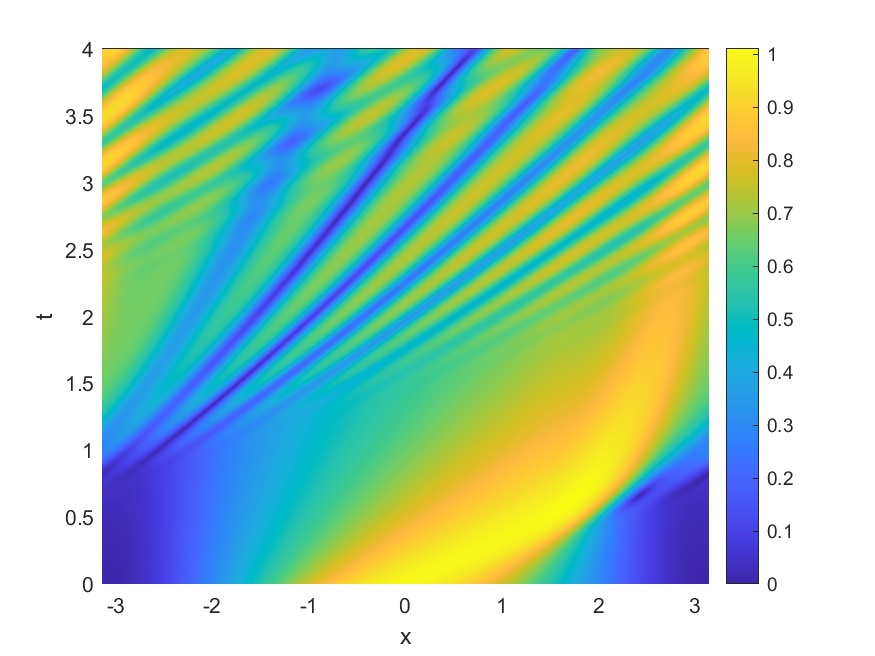}}
\centerline{\includegraphics[scale=.55]{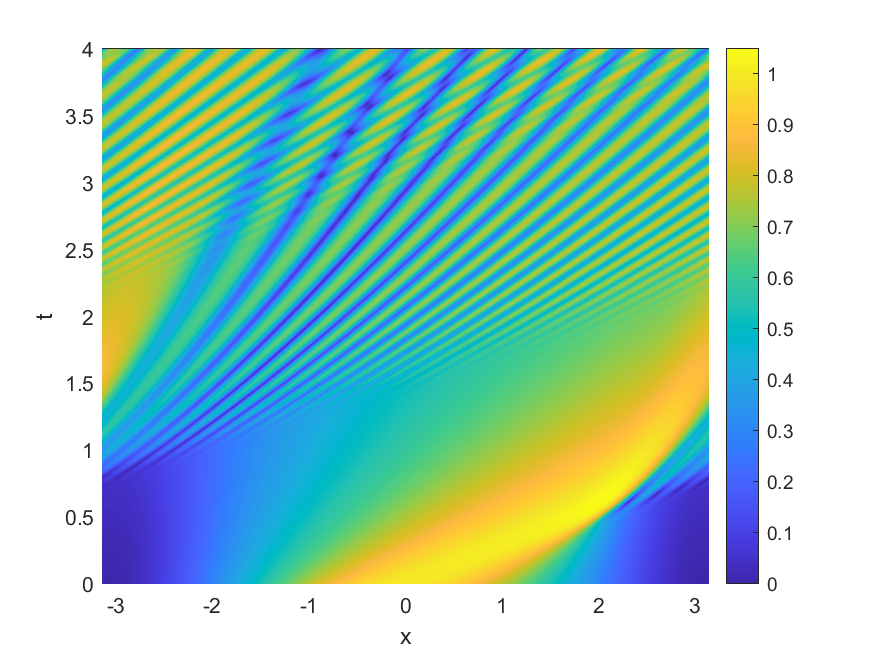}
  \includegraphics[scale=.55]{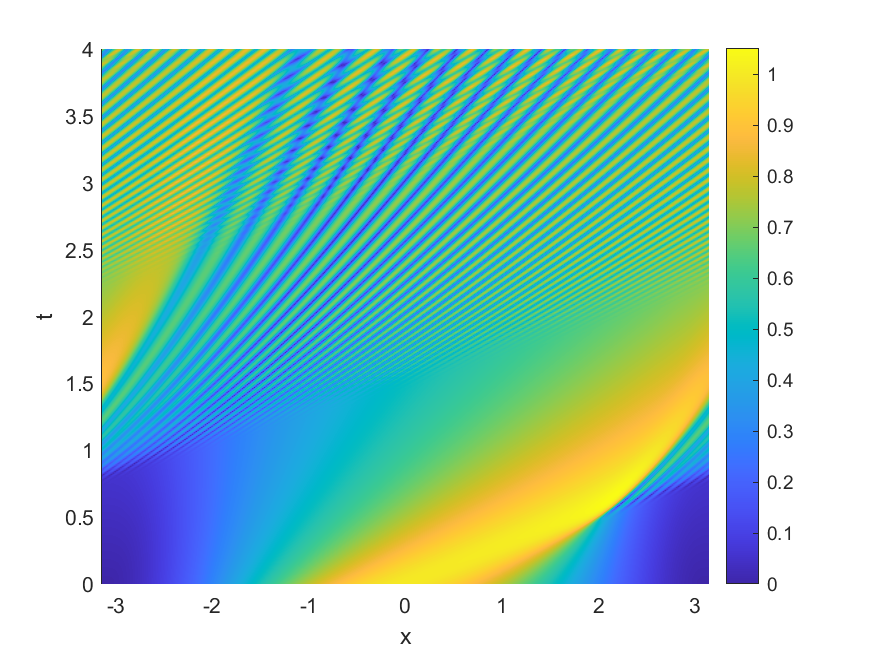}}
\caption{Density plots of the amplitude $|q(x,t)|$ of the solution of the DNLS equation with the ``raised cosine" IC for decreasing values of $\epsilon$. Top left: $\epsilon=0.2$. Top right: $\epsilon=0.1$. Bottom left: $\epsilon=0.05$. Bottom right: $\epsilon=0.025$.
See figure~\ref{fig7} for $\epsilon=0.01$.
}
\label{fig6}
\end{figure}

\begin{figure}[t!]
\centerline{\includegraphics[scale=.55]{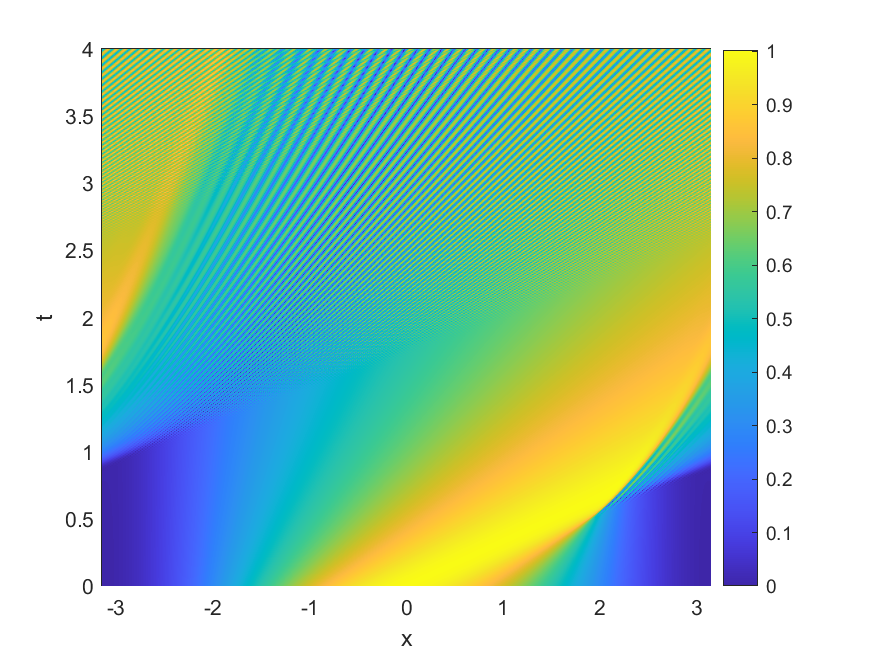}
  \includegraphics[scale=.55]{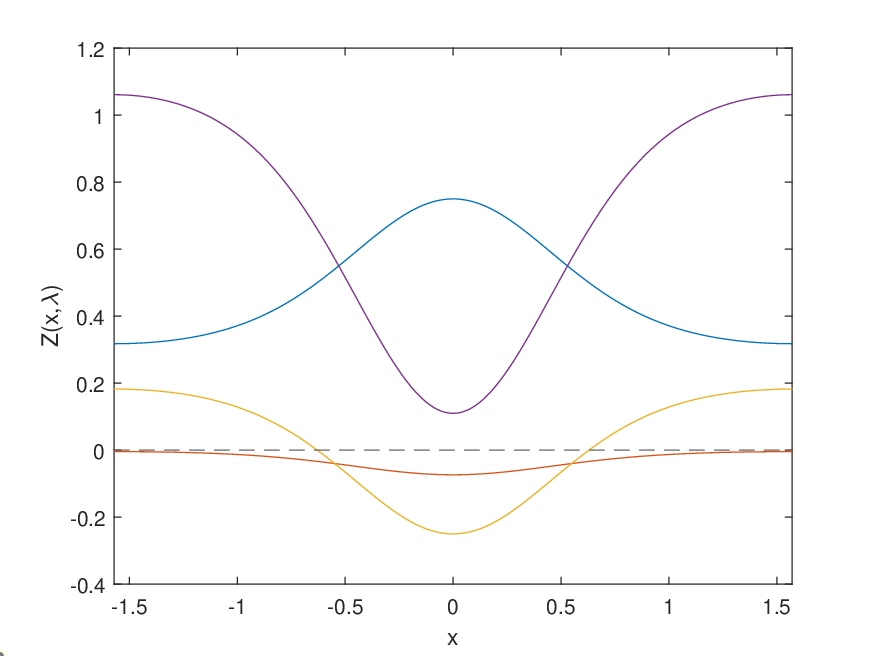}}
    \caption{Left: Density plot of the solution produced by the raised cosine potential with $\epsilon=0.01$.
    Comparing this plot and those in figure~\ref{fig6}, note how the spatial period of the oscillations in the dispersive regularization decreases as $\epsilon$ decreases. 
    Right: $Z(x,\lambda)=\lambda(\lambda+q^2(x))$ for the potential $q(x)=\e^{-\sin^2(x)}$ and various values of $\lambda$ corresponding to the four ranges described in section~\ref{s:asymptotics}. 
    Specifically:
    $\lambda=0.5$, in blue, corresponds to range~(i); 
    $\lambda=-1.1$, in purple, corresponds to range~(ii); 
    $\lambda=-0.5$, in yellow, corresponds to range~(iii); 
    $\lambda=-0.08$, in orange, corresponds to range~(iv). 
    Notice how the curve for $\lambda$ in range~(iii) is the only one whose sign changes from positive to negative, highlighting that range~(iii) is the only one with turning points.}
    \label{fig7}
\end{figure}

\section{Numerical Methods}
\label{s:numerics}

In this section we provide details of the methods used to numerically integrate the DNLS equation in semiclassical scaling~\eqref{e:dnls-eps}, 
as well as on the numerical computation of the Lax spectrum and evaluation on the monodromy matrix, 
the results of all which were presented in the previous sections.


\subsection{Numerical Solution of the DNLS Equation}

All numerical simulations of the DNLS equation~\eqref{e:dnls-eps} described in this work were performed using a pseudo-spectral method in space. 
The spatial derivatives in~\eqref{e:dnls-eps} can be written as
\[
q_{xx} = \F^{-1}_x[-k^2\F_k[q]],\qquad
(|q|^2q)_x = \F^{-1}_x[ik\F_k[|q|^2q],
\]
where $\F_k$ and $\F^{-1}_x$ denote respectively the direct and inverse Fourier transform operators, 
and $k$ is the wave number in the Fourier domain.
These operations were numerically computed by discretizing the spatial domain into $N$ evenly spaced grid points between $[-L,L]$, 
where $2L$ is the spatial period of the solution, as in section~\ref{s:ICs},
and by approximating the direct and inverse Fourier transform operators with the direct and inverse Fast Fourier Transforms.


A fourth-order Runge-Kutta method was then employed to evolve the solution in time. 
Rewriting~\eqref{e:dnls-eps} formally as $q_t = \mu(x,t) q$,  
the numerical integration step size $\Delta t$ was chosen such that all values of $\mu\,\Delta t$ lie within the stability region of the Runge-Kutta method. 
The resulting time step was the same for each spatial grid point, 
but it could change after each temporal step.


For all the plots in figures~\ref{fig1} and~\ref{fig6}, the number of Fourier modes used was $N=2^{10}$, 
the value of $L$ was chosen accordingly to the spatial period of each initial condition considered, 
and the solution was sampled at each $t_n = 10^{-2}n$. 
(Since the temporal step size can vary with each time step, whenever the largest safe time was larger than 
the difference between two consecutive sampling points, a smaller step was taken, in order to ensure that the solution slices were sampled over a uniform temporal grid.) 
Importantly, we should note that, owing to the scaling symmetry of equation~\eqref{e:dnls-eps}, 
the characteristic spatial periodicity of the individual locally periodic oscillations of the solution is $O(\epsilon)$, as is also easily seen in figure~\ref{fig6}.
Accordingly, as $\epsilon$ decreases, more Fourier modes are required in order to accurately capture the behavior
of solutions. 
For comparison purposes, figure~\ref{fig7} shows a plot of the solution with $\epsilon=0.01$. 
To properly capture the oscillations, $N=10^{12}$ Fourier modes were required in this case.

\subsection{Numerical Calculation of the Lax Spectrum via Floquet-Hill's Method}

When the potential matrix $Q(x,t)$ in the first half of the Lax pair \eqref{e:firstLax} is spatially periodic, the entire right-hand side of the equation is also periodic.
As a result, the Floquet-Hill method \cite{DeconinckKutz} can be used to calculate the Lax spectrum, as we show next. 
Recall that Floquet-Bloch theory allows us to rewrite the bounded solutions of \eqref{e:firstLax} as
\begin{equation}
\label{e:expper}
    \phi(x,\zeta)=\e^{\i\nu x}w(x,\zeta),
\end{equation}
where $w(x+2L,\zeta)=w(x,\zeta)$ and $\nu\in[0,\pi/L).$ Inserting \eqref{e:expper} into\eqref{e:firstLax} yields
\begin{equation}
\label{e:periodic}
    \left[-\zeta^2I-\i\zeta\sigma_3Q+\epsilon(i\partial_x-\nu I)\right]w=0.
\end{equation}
This equation is equivalent to \eqref{e:firstLax}, but with the main difference that $w(x,\zeta)$ is now also spatially periodic, which allows us to expand it in a Fourier series, obtaining
\begin{equation}
    \left[\zeta^2A+\zeta B+C_\nu^\epsilon\right]\hat{w}=0
\label{e:matrixeigenvalueproblem}
\end{equation}
with $\hat{w}=(\cdots,\hat{w}_{-1},\hat{w}_0,\hat{w}_1,\cdots)^T$, 
where $\hat{w}_j$ is the $j$-th Fourier coefficient of $w(x,\zeta)$, and
\begin{equation}
    A=\begin{bmatrix}
        -I & 0\\0 & -I
    \end{bmatrix},\quad
    B=\begin{bmatrix}
        0 & -i\mathcal{T}\\-i\mathcal{T}^*&0
    \end{bmatrix},\quad
    C_\nu^\epsilon=\begin{bmatrix}
        -\epsilon(K+\nu I) & 0 \\ 0 & \epsilon(K+\nu I)
    \end{bmatrix},
\end{equation}
Here $K=\diag(k_n)_{n\in\Integer}$ is the doubly infinite diagonal matrix of Fourier wavenumbers, with $k_n=n\pi/L$, and $\mathcal{T}$ is the doubly infinite Toeplitz matrix representing the convolution operator that arises from the Fourier series of $q(x)w(x,\zeta)$. 
The asterisk represents complex conjugation. 
In the case of a real potential, $\mathcal{T}=\mathcal{T}^*$. 

Selecting a fixed value of $\nu\in[0,\pi/L)$ and truncating equation~\eqref{e:matrixeigenvalueproblem} down to a finite matrix allows one to numerically compute the eigenvalues $\zeta$ that compose the Floquet-Bloch spectrum corresponding to the specific choice of~$\nu$. 
By repeating this process over a large grid of values for $\nu$, 
one can take the union of all computed Floquet-Bloch spectra as a numerical approximation of the Lax spectrum,
as per equation~\eqref{e:LaxFloquetunion}.

The accuracy of the method is determined by the number of Fourier modes left after truncation and by the accuracy of the solver for the quadratic eigenvalue problem. 
The density of the resulting numerical Lax spectrum corresponds to the number of values of~$\nu$ used in order to compute the union of Floquet-Bloch spectra. 
To generate the plots of the Lax spectrum in figure~\ref{fig2}, $N=2^7$ Fourier modes and $10^4$ values of $\nu$ on the interval $[0,\pi/L])$ were used. 
To numerically calculate the band edges for the right plot of figure~\ref{fig5}, $N=2^8$ Fourier modes were used for additional accuracy.  
The two specific values of $\nu$ corresponding to the band edges of the Lax spectrum are $\nu=0$ and $\nu=\pi/2L$.

\subsection{Numerical Calculation of the Monodromy Matrix}

As described above, the Floquet-Hill method generates a Floquet-Bloch spectrum in the complex $\zeta$ plane for a specific value of~$\nu$. 
The Floquet multiplier $\mu$ associated to the given value of $\nu$ is easily determined, 
and the Floquet discriminant at these points in the complex $\zeta$ plane can be obtained. 
The resulting values can also be verified by numerically calculating the value of the Floquet discriminant 
via direct integration of the scattering problem, as discussed next.

From equation \eqref{e:monodromy}, we choose a fundamental matrix solution $\Phi(x,\zeta)$ normalized so that $\Phi(0,\zeta)=I$, where $I$ is the $2\times2$ identity matrix. 
Then the monodromy matrix can be obtained simply as $M(\zeta)=\Phi(2L,\zeta)$.
The matrix $\Phi(x,\zeta)$ was calculated 
using the two choices $(1,0)^T$ and $(0,1)^T$ as initial conditions
and integrating equation~\eqref{scattering} numerically 
with a fourth-order Runge-Kutta method, with step size $\Delta x\leq10^{-3}$. 
From there, the Floquet discriminant $\Delta(\zeta) = \half \tr M(\zeta)$ is obtained. 
The resulting values were used to validate the accuracy of the Floquet-Hill method. 
The resulting values were also employed in figure~\ref{fig4} to test the accuracy of trace obtained through the WKB expansion.

\section{Details of the WKB Expansion and Asymptotic Calculations}
\label{s:wkb}

In this section we provide the details of the calculations that lead to the results presented in section~\ref{s:asymptotics}.

\subsection{Eikonal and Transport Equations}

We look for an asymptotic representation of the solutions to equation \eqref{e:vODE} of the form
\begin{equation}
\label{e:eiktrans}
    v(x)=(A(x)+O(\epsilon))\,\e^{\i S(x)/\epsilon},\quad \epsilon\downarrow0.
\end{equation}
Note that in~\eqref{e:eiktrans} and below, the $\lambda$ will often be suppressed for brevity. 
Substituting \eqref{e:eiktrans} into \eqref{e:vODE} yields the eikonal and transport equations respectively as
\bse
\label{e:eikonaltransport}
\begin{gather}
\label{eik}
    (S'(x))^2=Z(x,\lambda)\,,
\\
2S'(x)A'(x)+S''(x)A(x)+\sqrt{\lambda}q'(x)A(x)=0\,,
\end{gather}
\ese
where once again $Z(x,\lambda)=\lambda(\lambda-q^2(x))$ (recall $\lambda=\zeta^2$).
These equations can easily be integrated analytically if the sign of $Z(x,\lambda)$ is the same over the entire interval $[-L,L]$. 
For certain ranges of $\lambda$, however the sign of $Z(x,\lambda)$ changes at turning points, 
and the process is more complicated. 
As such, we will examine the solutions to equation~\eqref{e:vODE} of the form~\eqref{e:eiktrans} across several ranges of value $\lambda$, as determined by $q_{\max}^2$ and $q_{\min}^2$, the maximum and minimum values of the square of the potential.

\paragraph{Range (i): $\lambda<-q_{max}^2$.}
For all $\lambda$ in this range we have $Z(x,\lambda)>0$. and the resulting leading order WKB approximations are of the form
\begin{equation}
\label{e:vpos}
    v_\pm(x) = A_\pm(x)\e^{\i S_\pm(x)/\epsilon},
\end{equation}
with
\bse
\begin{gather}
\label{e:spos}
    S_\pm(x)=\pm\int_{-L}^x\sqrt{Z(x,\lambda)}\,\d x,
\\
\label{e:apos}
    A_\pm(x)=\frac{\sqrt{\mp\sqrt{|Z(x,\lambda)|}+\sqrt{\lambda}q(x)}}{\sqrt[4]{|Z(x,\lambda)|}}.
\end{gather}
\ese

\paragraph{Range (ii): $\lambda>0$.}
For all $\lambda$ in this range, we again have $Z(x,\lambda)>0$, and the leading order WKB approximation is the same as in region~(i).

\paragraph{Range (iii): $-q_{\max}^2<\lambda<-q_{\min}^2$.}
For $\lambda$ in this range and for a positive symmetric single-lobe potential, $Z(x,\lambda)$ has two real zeros at $x=\pm p(\lambda)$, i.e.,
$Z(\pm p(\lambda),\lambda)=0$.
The presence of zeros is the presence of turning points and the integration of the eikonal and transport equation becomes more complex and is discussed in detail in section~\ref{s:transitions}.

\paragraph{Range (iv): $-q_{\min}^2<\lambda<0$.}
Recall that this range is empty for any potential for which $q_{\min}^2=0$. 
For potentials for which this range is non-empty, one has that $Z(x,\lambda)<0$ for all $x$. 
Thus, in this region, the leading order WKB approximations are of the form 
\begin{equation}
\label{e:vneg}
    v_\pm(x) = A_\pm \e^{S_\mp(x)/\epsilon}\,,
\end{equation}
with
\bse
\begin{gather}
\label{e:sneg}
    S_\pm(x)=\pm\int_{-L}^x\sqrt{|Z(x,\lambda)|}\,\d x\,,
\\
\label{e:aneg}
    A_\pm(x)=\frac{\sqrt{\mp \i\sqrt{|Z(x,\lambda)|}+\sqrt{\lambda}q(x)}}{\sqrt[4]{|Z(x,\lambda)|}}\,.
\end{gather}
\ese

\subsection{Solution Formulae for Range~(iii)}
\label{s:transitions}

As stated above, when the value of~$\lambda$ is in range~(iii) there are turning points at $x = \pm p$. 
This splits the domain $[-L,L]$ into three regions, as well as a pair of transition regions. 
The precise location of these regions is as follows:
\begin{enumerate}
\advance\itemsep -6pt
    \item Region~1: $x\in[-L,-p)$.
    \item Transition~1: $x\in(-p-\delta,-p+\delta)$, $\delta>0$.
    \item Region~2: $x\in(-p,p)$.
    \item Transition~2: $x\in(p-\delta,p+\delta)$, $\delta>0$.
    \item Region~3: $x\in(p,L]$.
\end{enumerate}
Next we present the solution of equations~\eqref{e:eikonaltransport} in each region.
Then in section~\ref{s:connection} we discuss how one can patch the resulting expressions to obtain an approximation for the solution valid over the whole range~$[-L,L]$.

\paragraph{Region~1.}
In region 1, $Z(x,\lambda)>0$, implying that the WKB approximation for the general solution of the scattering problem of the form~\eqref{e:eiktrans} is
\vspace*{-1ex}
\begin{equation}
    v_1(x)=a_1^+v_+(x)+a_1^-v_-(x),
\label{e:v1}
\end{equation}
with $v_\pm(x)$ defined by \eqref{e:vpos}, $A(x)$ defined by \eqref{e:apos}, and $S(x)$ defined by \eqref{e:spos} with the lower integration limit replaced by $-p$ to avoid issues related to the sign change in $Z(x,\lambda)$.

\paragraph{Transition~1.}
The first transition is the neighborhood of the turning point $x=-p$. 
In this neighborhood, $Z(x,\lambda)=-a(x+p)+o(1)$ as $x\rightarrow-p$, with $a>0$, 
and the general solution to equation~\eqref{e:vODE} can be expressed as
\begin{equation}
    v_{1\rightarrow2}(x)=c_1^-\text{Ai}[\xi(x,\lambda)]+c_1^+\text{Bi}[\xi(x,\lambda)],
\end{equation}
where $\xi(x,\lambda)=a^{1/3}(x+p)/\epsilon^{2/3}$ and Ai($\cdot)$ and Bi($\cdot)$ are the Airy functions \cite{NIST}.

\paragraph{Region~2.}
There are two different but equivalent representations of the WKB approximation for the solution of equation~\eqref{e:vODE} in this region, depending on the starting point of integration. 
Since $Z(x,\lambda)<0$ here, these representations are of the form:
\bse
\begin{gather}
    v_2(x)=a_2^+v_+(x)+a_2^-v_-(x)\,,
    \label{e:v2}
\\
    \bar{v}_2(x)=\bar{a}_2^+\bar{v}_+(x)+\bar{a}_2^-\bar{v}_-(x)\,,
\end{gather}
\ese
where
\vspace*{-1ex}
\bse
\begin{gather}
\label{e:vnobar}
    v_\pm(x)=A_\pm(x)\exp\left(\mp\int_{-p}^x\sqrt{|Z(s,\lambda)|}\,\d s/\epsilon\right)
\\
\label{e:vbar}    
    \bar{v}_\pm(x)=A_\pm(x)\exp\left(\mp\int_{p}^x\sqrt{|Z(s,\lambda)|}\,\d s/\epsilon\right),
\end{gather}
\ese
with $A_\pm(x)$ given by \eqref{e:aneg}, and where the integrals in the exponentials in equations~\eqref{e:vnobar} and~\eqref{e:vbar} are the same as \eqref{e:sneg}, except that the lower integration limit has been replaced by $-p$ and $p$ respectively.

\paragraph{Transition 2.}
The second transition region is similar to the first, with the difference being that now $Z(x,\lambda)=b(x-p)+o(1)$ as $x\rightarrow p$, with $b>0$ (notice the lack of the negative sign). 
Similarly to transition~1, the solution of equation~\eqref{e:vODE} here is
\begin{equation}
    v_{2\rightarrow3}=c_2^-\text{Ai}[\eta(x,\lambda)]+c_2^+\text{Bi}[\eta(x,\lambda)]
\end{equation}
where $\eta(x,\lambda)=-b^{1/3}(x-p)/\epsilon^{2/3}$.

\paragraph{Region~3.}
In region 3 $Z(x,\lambda)>0$ like region 1 and the WKB solution of \eqref{e:vODE} is similarly of the form
\begin{equation}
    v_3(x)=a_3^+v_+(x)+a_3^-v_-(x),
\end{equation}
with $v_\pm(x)$ defined by \eqref{e:vpos}, $A(x)$ defined by \eqref{e:apos}, and $S(x)$ defined by \eqref{e:spos} with the lower integration limit replaced by $p.$

\subsection{Connection Matrices}
\label{s:connection}

For the purpose of calculating the monodromy matrix, 
one needs an explicit representation for the solution over the whole range $[-L,L]$.
Accordingly, we now asymptotically match the expressions for the solution in each region to the nearby transition regions, and we obtain a connection between the two expressions for the solution in region~2.

As region~1 approaches the first transition region from the right, the two linearly independent portions of the leading order WKB solution~\eqref{e:v1} in region~1 become
\begin{equation}
    v_\pm(x)=\frac{\sqrt{|\lambda|}}{\sqrt[4]{a|x+p|}}\e^{\mp\frac{2}{3} \i a^{1/2}(x+p)^{3/2}/\epsilon}\quad x\rightarrow-p^-.
\end{equation}
Using the known asymptotic expansions of the Airy functions, we can then match the resulting expansion for $v_1(x)$ as $x\rightarrow-p^-$ to that of $v_{1\rightarrow2}(x)$ as $\xi\rightarrow-\infty$, and thereby obtain the first connection formula as
\begin{equation}
    \begin{bmatrix}
        c_1^-\\c_1^+
    \end{bmatrix}=C_1
    \begin{bmatrix}
        a_1^-\\a_1^+
    \end{bmatrix},\quad
    C_1=\frac{(\pi|\lambda|)^{1/2}}{(a\epsilon)^{1/6}}\begin{bmatrix}
        \e^{\i\pi/4}&\e^{-\i\pi/4}\\\e^{-\i\pi/4} & \e^{\i\pi/4}
    \end{bmatrix}.
\end{equation}
Next, as region~2 approaches the first transition region from the right, the two linearly independent portions of the leading order WKB solution~\eqref{e:v2} in region~2 become
\begin{equation}
    v_\pm(x)=\frac{\sqrt{|\lambda|}}{\sqrt[4]{a(x+p)}}\e^{\pm \frac{2}{3}a^{1/2}|x+p|^{3/2}/\epsilon}.
\end{equation}
Using the asymptotic expansions of the Airy functions once again, this time at $\xi\rightarrow\infty$, 
we can match the resulting expansion for $v_2(x)$ as $x\rightarrow-p^+$, to obtain the second connection formula as
\begin{equation}
    \begin{bmatrix}
        a_2^-\\a_2^+
    \end{bmatrix}=C_2
    \begin{bmatrix}
        c_1^-\\c_1^+
    \end{bmatrix},\quad
    C_2=\frac{(a\epsilon)^{1/6}}{(\pi|\lambda|)^{1/2}}
    \begin{bmatrix}
        0&1\\1/2&0
    \end{bmatrix}.
\end{equation}
Next, a relation between the coefficients of the two representations of the WKB solution in region 3 is needed. This third connection formula is
\begin{equation}
    \begin{bmatrix}
        \bar{a}_2^-\\\bar{a}_2^+
    \end{bmatrix}=C_3
    \begin{bmatrix}
        a_2^-\\a_2^+
    \end{bmatrix},\quad
    C_3=\e^{\sigma_3\int_{-p}^p\sqrt{|Z(s,\lambda)|}dx/\epsilon}
\end{equation}
where $\sigma_3=\diag(1,-1)$.
Next, as region~2 approaches the second turning point from the right, the two linearly independent portions of the leading order WKB solution have the expansion
\begin{equation}
    v_{\pm}(x)=\frac{\sqrt{|\lambda|}}{\sqrt[4]{b|x-p|}}\e^{\pm\frac{2}{3}b^{1/2}|x-p|^{3/2}/\epsilon}\quad x\rightarrow p^+\,.
\end{equation}
Matching these to the expansions for the Airy functions in transition~2 as $\eta\rightarrow\infty$ yields the fourth connection formula as
\begin{equation}
    \begin{bmatrix}
        c_2^-\\c_2^+
    \end{bmatrix}=C_4
    \begin{bmatrix}
        \bar{a}_2^-\\\bar{a}_2^+
    \end{bmatrix},\quad C_4=
    \frac{(\pi|\lambda|)^{1/2}}{(b\epsilon)^{1/6}}
    \begin{bmatrix}
        2&0\\0&1
    \end{bmatrix}\,.
\end{equation}
The final connection formula is obtained when approaching the second turning point from the right. 
In region~3, the leading order WKB solution is obtained by
\begin{equation}
    \begin{bmatrix}
        a_3^-\\a_3^+
    \end{bmatrix}=C_5
    \begin{bmatrix}
        c_2^-\\c_2^+
    \end{bmatrix},\quad
    C_5=\frac{(b\epsilon)^{1/6}}{2(\pi|\lambda|)^{1/2}}
    \begin{bmatrix}
        \e^{\i\pi/4}&\e^{-\i\pi/4}\\ \e^{-\i\pi/4}&\e^{\i\pi/4}
    \end{bmatrix}.
\end{equation}
Lastly, we can compute the overall connection formula relating the coefficients at one end of the spatial period to the other as $C=C_5C_4C_3C_2C_1$ explicitly calculated as
\begin{equation}
    C=\begin{bmatrix}
        \cosh(S_1(\lambda)/\epsilon+\ln(2))&\i\sinh(S_1(\lambda)/\epsilon+\ln(2))\\
        -\i\sinh(S_1(\lambda)/\epsilon+\ln(2))&\cosh(S_1(\lambda)/\epsilon+\ln(2))
    \end{bmatrix}
\end{equation}
with $S_1(\lambda)$ as in equation~\eqref{e:S1S2def}.

\subsection{Calculation of the Floquet Discriminant}

We now have all the necessary pieces to calculate the Floquet discriminant in each range of values of~$\lambda$. Using equation~\eqref{e:monodromy} evaluated at $x=0$, we can compute the monodromy matrix as
\begin{equation}
    M(\lambda)=\Phi(-L,\lambda)^{-1}\Phi(L,\lambda).
\end{equation}
Note that the periodicity of the potential means $Z(-L,\lambda)=Z(L,\lambda)$ and $A_\pm(-L)=A_\pm(L)$ where $A(x)$ is defined for each range of $\lambda$. 
It will also be convenient to define
\begin{equation}
\label{e:matrixcoef}
    \Phi_o=\begin{bmatrix}
        A_-(L)&A_+(L)\\
        A_-(L)\sqrt{|Z(L,\lambda)|}/\epsilon & -A_+(L)\sqrt{|Z(L,\lambda)|}/\epsilon
    \end{bmatrix}
\end{equation}
where the particular definition for~$A_\pm(L)$ depends on the range that~$\lambda$ is in.

\paragraph{Ranges~(i) and~(ii).}
The WKB solutions in ranges~(i) and~(ii) have the same form, so the calculation of the monodromy matrix is the same. At leading order, we have the following values for the fundamental solution matrix in ranges~(i) and~(ii):
\vspace*{-1ex}
\begin{gather}
    \Phi(-L,\lambda)=\Phi_o\,,
\qquad
    \Phi(L,\lambda)=\Phi_o\,\e^{\i\sigma_3\int_{-L}^L\sqrt{Z(x,\lambda)}\d x/\epsilon}\,.
\end{gather}
Using equation \eqref{e:monodromy}, we can then easily obtain the monodromy matrix, which in turn yields
the Floquet discriminant in equation~\eqref{e:Deltacosine}.

\paragraph{Range~(iii).}
For an even single-lobe potential, the even symmetry gives
\begin{equation}
    \int_{-L}^{-p(\lambda)}\sqrt{|Z(x,\lambda)|}\d x=\int_{p(\lambda)}^L\sqrt{|Z(x,\lambda)|}\d x=S_2(\lambda).
\end{equation}
This allows us to obtain the following pair of values for the fundamental matrix solution out to leading order:
\begin{equation}
    \Phi(-L,\lambda)=\Phi_o\e^{-S_2(\lambda)\sigma_3/\epsilon}\,,
\qquad
    \Phi(L,\lambda)=\Phi_o\e^{S_2(\lambda)\sigma_3/\epsilon}\,C\,,
\end{equation}
where $C$ is the connection matrix calculated in section~\ref{s:connection}. 
Using equation~\eqref{e:monodromy} to obtain the monodromy matrix and taking the trace yields
the expression in equation~\eqref{e:trM3}.

\paragraph{Range~(iv).}
Range (iv) has no turning points, so the calculation of the monodromy matrix is similar to that for ranges~(i) and~(ii). 
Namely, the values of the fundamental matrix solution in range~(iv) are
\begin{equation}
    \Phi(-L,\lambda)=\Phi_o\,,
\qquad
    \Phi(L,\lambda)=\Phi_o\e^{-\sigma_3\int_{-L}^L\sqrt{|Z(x,\lambda)|}\d x/\epsilon}\,,
\end{equation}
and the resulting trace of the monodromy matrix is given in equation~\eqref{e:trM4}.

\section{Concluding remarks}
\label{s:conclusions}

In summary, we studied the the semiclassical limit of the DNLS equation with real periodic ICs by examining the behavior of single-lobe potentials. 
Numerical evaluation of the Lax spectrum in the semiclassical limit suggest that the spectrum of single-lobe potentials clusters to the real and imaginary axes of the spectral variable~$\zeta$ as $\epsilon\downarrow0$. 
Motivated by this result, we introduced an auxiliary spectral variable $\lambda=\zeta^2$ and derived an asymptotic characterization of the behavior of the Lax spectrum as a function of real values of $\lambda$ as $\epsilon\downarrow0$. 
This characterization divides the real line into four distinct regions with boundaries at 0, $-q_{\text{min}}^2$, and $-q_{\text{max}}^2$. 
The ranges $(-\infty,-q_{\text{max}}^2)$ and $(0,\infty)$ will always be spectrum, reminiscent of the continuous spectrum of the DNLS with nonvanishing boundary conditions on the entire line. The range $(-q_{\text{max}}^2,-q_{\text{min}}^2)$ will have spectral bands that shrink as $\epsilon$ decreases and $(-q_{\text{min}}^2,0)$ will always be a spectral gap. 
We then showed that the number of bands in the range $(-q_{\text{max}}^2,-q_{\text{min}}^2)$ scales like $1/\epsilon$. Furthermore, we showed that if $\kappa$ is defined as the maximum relative length for a spectral band to be treated as a discrete point, all spectral bands in the range $(-q_{\text{max}}^2,-q_{\text{min}}^2)$ will effectively be discrete points as $\epsilon\downarrow0$ regardless of the choice of $\kappa$. As such, the spectrum of the DNLS for single-lobe potentials in the semiclassical limit is similar to the spectrum of the DNLS on the real line with nonvanishing boundaries equal to $q_{\text{max}}$ and an infinite number of points in the discrete spectrum located on the imaginary axis in the range $(\i q_{\text{max}},-\i q_{\text{min}})$.

A natural comparison to the present study of the DNLS equation in the semiclassical limit is the focusing NLS equation in the semiclassical limit, 
namely the PDE $i\epsilon q_t + \epsilon^2 q_{xx} + 2|q|^2q = 0$,
results for which can be found in \cite{BO2020}. 
In terms of the modified spectral variable $\lambda = \zeta^2$, the Lax spectrum for both equations contains the entire range $(0,\infty)$ as a spectral band. 
The results for $\lambda<0$, however, are almost reversed, 
with the focusing NLS having a spectral band for $\lambda$ in the range $(-q_{\text{min}}^2,0)$ and no spectral band in the range $(-\infty,-q_{\text{max}}^2)$,
in contrast to the results in section~\ref{s:asymptotics}.
This difference can be easily understood by looking at the asymptotic characterization of the spectral problem for each equation and by looking at the difference in the corresponding eikonal equations. 
Recall that the eikonal equation for the focusing NLS equation is  $(S')^2=\lambda-q^2(x)$, 
which is almost the same as equation~\eqref{eik} for the DNLS equation, 
with the main difference being that the focusing NLS equation does not have an extra factor of~$\lambda$ on the right-hand side. 
This additional factor in the DNLS equation flips the sign of the right-hand side. 
Remembering that the WKB solution has exponential or oscillatory behavior depending on whether the right-hand side of the eikonal equation is imaginary or real, respectively, it should now be easy to see how this sign change results in the opposite behavior for the DNLS equation compared to the NLS equation.

Another major difference between the focusing NLS equation and the DNLS equation is the kind of solutions generated by the resulting spectra. 
In the focusing NLS equation, the effective discrete eigenvalues of the Lax spectrum correspond to solitons that all have zero velocity. This leads to a soliton condensate that is essentially confined to a single spatial period. 
This also means that periodic initial conditions and localized initial conditions produce similar solutions. 
In contrast, the effective discrete discrete eigenvalues of the Lax spectrum correspond to solitons that have a nonzero velocity. 
As a result, faster moving solitons will wrap around a spatial period and will eventually catch up to slower moving solitons, similarly to what happens in the corresponding problem for the KdV equation~\cite{ZabuskyKruskal}.

We believe that the results of this work are foundational, in particular since the spectral analysis applies equally well to all equations in the Kaup-Newell hierarchy. 
We also note that the DNLS equation can be mapped into the modified NLS equation and the Gerdjkov-Ivanov equation~\cite{Clarkson1987}, both of which have also been studied extensively.
Therefore, it is likely that the present analysis can also be used to study the semiclassical limit of those equations.

We also believe that the results of this work open up several interesting follow-up problems.  
For example, an immediate question that should be investigated involves a precise identification of the precise spectral content of the individual solitons generated in the break-up of the initial pulse.
Another project will be the use of the recently derived Whitham modulation equations for the DNLS equation~\cite{Kamchatnov2018} in order to obtain a quantitative description of the dispersive shock waves resulting from the breakup of the initial conditions, similarly to what was done for the Kortweg-de\,Vries and the focusing and defocusing NLS equation in~\cite{GurevichPitaevski1974,PHYSD1995v87p186,SJAM59p2162,BK2006,HA2007,B2018}.
Yet another problem will be a rigorous description of the resulting phenomena using suitable asymptotics on the inverse problem of the IST, similarly to what was done in \cite{BuckinghamVenakides,Jenkins2015,BM2017} for the focusing and defocusing NLS equations.

Finally, an interesting application will be the study of soliton gases for the DNLS equation, 
following on the recent results of~\cite{Zhong2025}.
For the focusing NLS equation, it was shown in \cite{PRE2025} that the addition of a small amount of noise to a semiclassical potential
serves to effectively randomize the phases of the nonlinear excitations thanks to the
presence of modulational instability, resulting in the formation of a fully developed soliton gas
(or, more generally, a breather gas).
The resulting structures were then recently observed experimentally in \cite{OL2025}.
It would therefore be interesting to explore whether similar results apply for the DNLS equation.
We hope that the present work and the above discussion will stimulate further study on some of these problems in the near future.

\medskip
\let\em=\it

\makeatletter
\def\@biblabel#1{#1.}
\def\doibase{http://dx.doi.org/}
\def\reftitle#1{``#1''}
\def\booktitle#1{\textit{#1}}
\def\href#1#2{#2}
\def\journal#1#2{#1 {\bf#2}}
\makeatother
\small


\end{document}